\documentclass[12pt]{article}
\usepackage{amsmath}
\usepackage{graphicx,psfrag,epsf}
\usepackage{enumerate}
\usepackage{natbib}
\usepackage{ulem}

\usepackage{cleveref}
\usepackage{color}
\usepackage{amssymb}
\usepackage[abs]{overpic}

\newcommand{\blind}{0}

\newcommand{\red}[1]{\textcolor{red}{#1}}

\newcommand{\mc}[1]{\mathcal{#1}}
\newcommand{\mb}[1]{\mathbb{#1}}
\newcommand{\mf}[1]{\mathbf{#1}}
\newcommand{\bs}[1]{\boldsymbol{#1}}
\newcommand{\mcI}{\mathcal{I}}

\newif\ifplots
\plotstrue 
\begin{document}

\bibliographystyle{apalike}

\def\spacingset#1{\renewcommand{\baselinestretch}%
{#1}\small\normalsize} \spacingset{1}

%%%%%%%%%%%%%%%%%%%%%%%%%%%%%%%%%%%%%%%%%%%%%%%%%%%%%%%%%%%%%%%%%%%%%%%%%%%%%%

\if0\blind
{
 \title{\bf Guided projections for analysing the structure of high-dimensional data}
  \author{Thomas Ortner\thanks{This work has been partly funded by the Vienna Science and Technology Fund (WWTF) through project ICT12-010 and by the K-project DEXHELPP through COMET - Competence Centers for Excellent Technologies, supported by BMVIT, BMWFW and the province Vienna. The COMET program is administrated by FFG.}\\
    Institute of Statistics and Mathematical Methods in Economics, \\
    Vienna University of Technology \\
    and \\
    Peter Filzmoser \\
    Institute of Statistics and Mathematical Methods in Economics, \\
    Vienna University of Technology \\
     and \\
    Maia Zaharieva \\
     Institute of Software Technology and Interactive Systems,  \\
    Vienna University of Technology  \\
    and \\
    Christian Breiteneder\\
    Institute of Software Technology and Interactive Systems,  \\
    Vienna University of Technology  \\
    and \\ 
    Sarka Brodinova \\
    Institute of Software Technology and Interactive Systems,  \\
   Vienna University of Technology 
    }
  \maketitle
} \fi

\if1\blind
{
  \bigskip
  \bigskip
  \bigskip
  \begin{center}
    {\LARGE\bf Title}
\end{center}
  \medskip
} \fi

\bigskip
\begin{abstract}
A powerful data transformation method named guided projections is proposed creating new possibilities to reveal the group structure of high-dimensional data in the presence of noise variables. Utilising projections onto a space spanned by a selection of a small number of observations allows measuring the similarity of other observations to the selection based on orthogonal and score distances. Observations are iteratively exchanged from the selection creating a non-random sequence of projections which we call guided projections. 
In contrast to conventional projection pursuit methods, which typically
identify a low-dimensional projection revealing some interesting features contained
in the data, guided projections generate a series of projections that 
serve as a basis not just for diagnostic plots but to directly investigate the group structure in data.
Based on simulated data we identify the strengths and limitations of guided projections in comparison to commonly employed data transformation methods. We further show the relevance of the transformation by applying it to real-world data sets.
\end{abstract}

\noindent%

\spacingset{1.45}
% ===============================================================
\section{Introduction}\label{sec:introduction}
\label{sec:motivation}

%Classical data analysis is facing various challenges. 
One of the most frequent problems in classical data analysis is the high dimensionality of data sets. In this paper we propose a novel method for data transformations, called \textit{guided projections}, in order to reveal structure in high-dimensional, potentially flat data. 
The presented approach uses subsets of observations to locally describe the data structure close to the subsets and measures similarity of all observations to these subsets utilising the projection onto such subsets. Exchanging observations one by one, we continuously change the location of the local description.
By \textit{guiding} the way these subsets are selected, we receive a sequence of projections which can be directly used as a data transformation, as well as a method for visualising group structure in high-dimensional data. 
In this paper we present some theoretical background and properties of the proposed guided projections and focus on the general separation between groups in data and how this separation, measured by various validation indices, is affected by the transformation. 
Furthermore, we compare with existing methods and discuss the strengths and the limitations of guided projections in experiments on both synthetic and real-world data.

%Some of the most frequently occurring problems are the high-dimensionality and the size of available data sets. In many cases it is a-priori not known which information should be collected. Therefore, many applications simply collect more and more data. This leads to thousands or millions of variables describing objects, combined with a vast number of observations. The large number of observations leads to a computational burden which strongly limits the use of possible approaches even considering the technological development of the last decade. The high-dimensionality further leads to challenges when measuring distances between observations. After all, David Donoho's statement from 2000, that classical statistical methods are not designed to deal with a world awash in data still holds \citep{Donoho2000}.

Let $\bs{X} \in \mathbb{R}^{n \times p}$ denote a data matrix, with $p$ variables and $n$ observations. We further assume that some unknown group structure is present in the observations. In particular we want to consider the possibility that $p$ is larger than $n$. A large number of variables leads to two main problems we would like to address: First, the cost of computational effort for computing all pairwise distances is $O(n^2p)$. While we cannot directly influence $n$, a reduction in $p$ will directly affect computation time. Second, in general, not all $p$ variables hold relevant information about the underlying group structure \citep{Hung2003}. 
Assume that the data contain some inherent group structure.
In accordance to \cite{Hung2003} we call variables contributing to a group separation \textit{informative} and variables not contributing to a group separation \textit{non-informative} variables. Accordingly, let us assume $p=p_1 + p_2$, where $p_1$ denotes the number of informative variables, and $p_2$  denotes the number of non-informative variables. 
If $p_1$ increases, a dimension reduction can considerably reduce the 
computational burden. If, however, $p_2$ increases, the variance from non-informative variables will mask the separation provided from informative variables. One possible solution to deal with this masking effect is the application of a data transformation to reveal the group structure in a lower dimensional space. The analysis of effects of such data transformations is the focus of this paper.

\medskip
A variety of data transformations has been proposed in the past. We present a small selection of commonly employed methods before proposing a novel approach for data transformation. 

Classical \textit{variable selection} methods rely on selecting a subset of features which are useful for identifying group structures in data \citep{Guyon2003}. 
A dimension reduction to a small subset of variables, based on some statistic on the 
distribution of the variables usually provides a suboptimal framework for the analysis of present group structures. One example is the commonly applied method of selecting the 5\% of variables with the largest  variance for gene expression data. From the variance itself, in general, it can not be concluded whether or not variables are informative. 

With the focus on computation time, \textit{Random Projections} (RP) \citep{Achlioptas2003} randomly project $\bs X$ onto $\mathbb{R}^{n\times k}$, $k<p$, preserving the expected pairwise distances.
There are different ways to identify the required projection matrices. In this paper we use iid normally distributed coefficients as proposed in \cite{Li2006}. Such random projections always contain contributions in the same proportion from non-informative variables as from informative variables though. 

An approach different from random projections and variable selection is \textit{Principal Component Analysis} (PCA) \citep[e.g.][]{Abdi2010} which is likely the most studied data transformation method. PCA identifies $k<p$ linear combinations of variables, maximising the variances of each resulting component under the restriction of orthogonality. Such components are called principal components. Classical PCA is subject to restrictions like identifying linear subspaces only. Furthermore, the differences in distances remain masked, since the principal components contain an increasing portion of the non-informative variables with an increasing number of such variables. The problem of linearity has been addressed in several publications \citep{Gorban2008, Leeuw2011}. We will consider \textit{Diffusion maps} (DIFF) \citep{Coifman2006} as one possible modification, where PCA is performed on the transformed data, based on distances measured by random walk processes. We will further consider \textit{Sparse Principal Component Analysis} (SPC) \citep{Zou2005, Zou2006, Witten2009}, since the goal of sparse PCA is to avoid the second problem we addressed, namely the presence of non-informative variables, by downweighting the non-informative variables.

A more general projection approach is \textit{Projection Pursuit} \citep{Friedman1974} where a projection onto a low-dimensional subspace is identified, maximizing a measure of interest like non-normality. This approach can further be generalized to similarities between estimated and general density functions \citep{Cook1993} and visualised using so called guided tours \citep{Cook1995}. 
There are also proposals for modifications of the projection pursuit index in order
to cope with high-dimensional data \citep{Lee2010}.
With the main intension of visualisation and visual analysis of projections, the dimension of the projection pursuit is mostly limited between one and three.

\medskip
After performing such a data transformation, one hopes to yield more information about the underlying group structure of the data. Such information can be measured in terms of performance with respect to a subsequent application of
outlier detection methods, discriminant analysis, clustering methods, and other related methods. 

%In this paper we focus on the general separation between groups in data and how this separation, measured by validation indices, is affected by data transformations. 
%We present a novel approach for data transformation including some theoretical background and properties of this transformation.
%Furthermore, we compare with existing methods and discuss its strengths and limitations by using synthetic data and finally show the impact based on a real-world data set.

The paper is structured as follows. The methodology and properties of our approach is presented in Section \ref{sec:methodology} providing insight on the effects of the transformation as well as a possibility for diagnostic plots. We define synthetic setups for the comparison of the newly introduced method with existing data transformation methods in Section \ref{sec:simulation} and report the results of the performed comparison. In Section \ref{sec:fruit} we apply the methods to two real-world data sets to illustrate the relevance of guided projections. Finally, we provide conclusions and an outlook on possible extensions and applications of the proposed method in Section \ref{sec:conclusion}.

% ===============================================================
\section{Guided projections}\label{sec:methodology}

Let $\bs{X} \in \mathbb{R}^{n\times p}$ denote the data matrix to be analysed. We further assume, that the observations $\bs x_i$, $i \in \{ 1 \dots n \}$, are randomly drawn from one of the distributions $F_1, \dots, F_m$, $m<n$. Therefore, up to $m$ groups are present in our data structure. 

The basic concept of guided projections is to find a non-random series of projections providing directions where differences between occurring groups are present. 
Each projection will be described by a selection of observations spanning the projection space. 
Any such selection describes the data structure close to the selected observations. 
By using a small number of observations for the projection, we avoid the masking effects of outlying observations on the description. In this context an outlying observation is an observation which is likely to be from a different group. This concept is visualised in Figure \ref{fig:example2groups} for a two dimensional space, using Mahalanobis distances as a representative for the similarity between observations. 
Since we assume a high-dimensional flat data space, we describe the properties of observations with respect to each specific projection. Therefore we use two
distance measures described in \cite{Hubert05}, the \textit{orthogonal distance} and the \textit{score distance}. Using these distances, we iteratively identify a series of observations leading to the series of projections (guided projections).

\ifplots
\begin{figure}[!ht]
\centering
\includegraphics[width=0.8\linewidth]{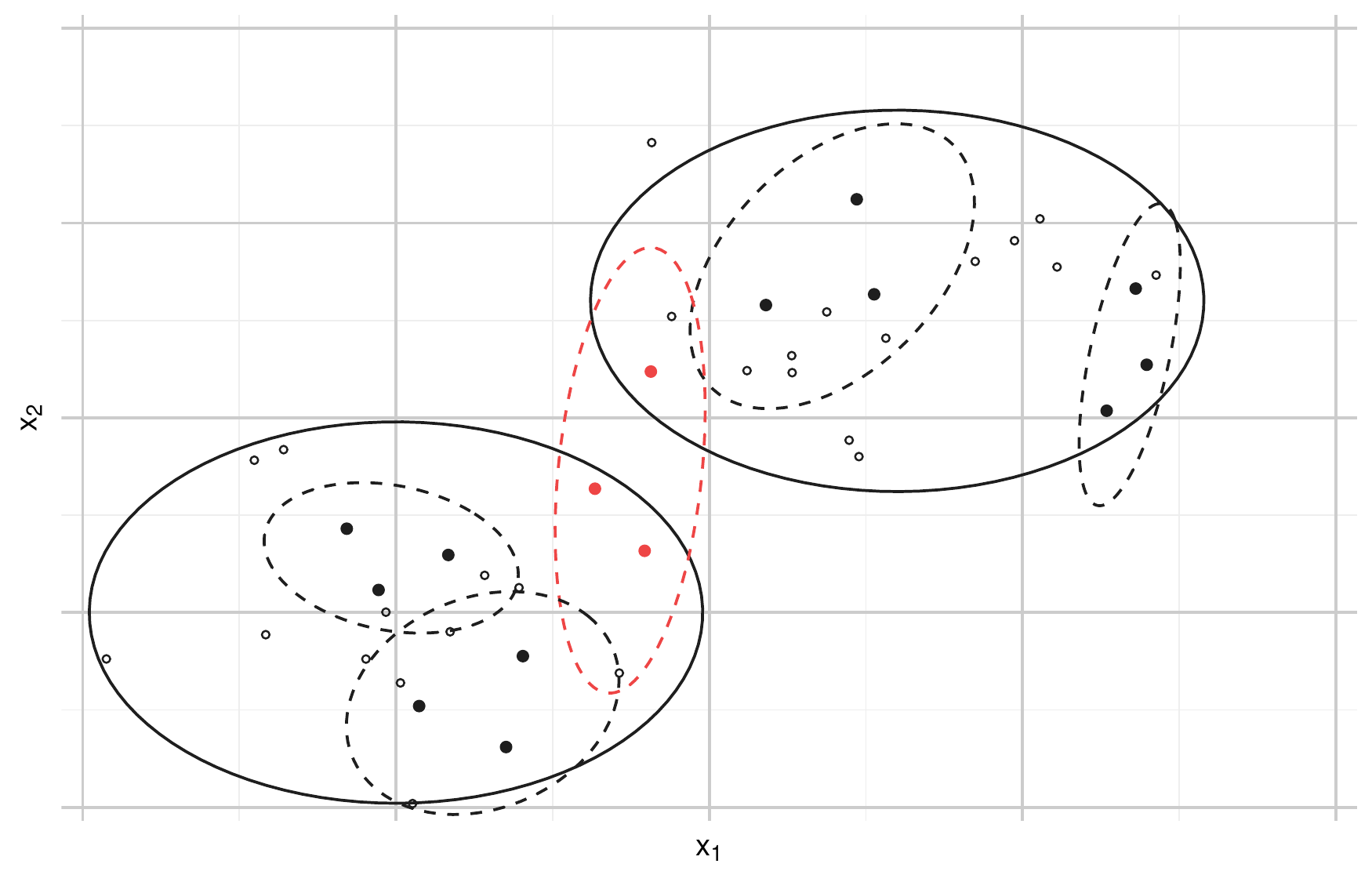}
\caption{This plot demonstrates the concept of guided projections. 
The figure shows two group structures and the corresponding true covariance structures described by solid ellipses. Each small subset of three observations, represented by solid points, will provide a local approximation of this group structure as visualised by the dashed ellipses. The aim of the proposed guided projections approach is to provide a series of such selections, offering a good overall description of the present group structures. Each black subset represents selections from the same group, providing useful information about the group separation, the red subset represents a mixed selection, where the group structure is masked, i.e. 
%. This can be seen based on 
observations from both groups are present inside the ellipse.
}
\label{fig:example2groups}
\end{figure}
\fi

\subsection{Orthogonal and score distances}

Let $\mathbb{P}$ denote the set of all orthogonal projections $P$ from $\mathbb{R}^p$ onto $\mathbb{R}^{{q-1}}$, where $p$  is the number of variables in the original space and ${q-1}$ the fixed dimension of the projected space. Each projection $P$ can be represented by its projection matrix $\bs V_P'$, where $\bs V_P \in \mathbb{R}^{p \times q-1}, P \in \mathbb{P}$: 
\begin{equation}
\forall P \in \mathbb{P}: \exists  \bs V_P \in \mathbb{R}^{p \times {q-1}}: P(\bs{x}) = \bs V_P'\bs x \hspace{15pt} \forall \bs x \in \mathbb{R}^p
\end{equation}

Given a projection $P \in \mathbb{P}$, we define the \textit{orthogonal distance} ($OD_P$) of an observation $\bs x \in \mathbb{R}^p$ to a projection space defined by $P$, given a location $\pmb \mu$ as 
\begin{equation}
OD_P(\bs x) = || \bs x -\bs{\mu} - V_PV_P'(\bs x-\bs{\mu}) ||, 
\label{od_classic}
\end{equation}
and the \textit{score distance} ($SD_P$) of $\bs x$, given the location $\bs \mu$ and the covariance matrix $\bs \Sigma_P$ of the distribution in the projection space as
\begin{equation}
SD_P(\bs x) = \sqrt{ (\bs V_P'(\bs x-{\bs \mu}))' {\bs \Sigma}_{P}^{-1} (\bs V_P'(\bs{x}-{\bs{\mu}}))},
\label{sd_classic}
\end{equation}
where $||.||$ stands for the Euclidean norm.

This definition slightly differs from the original concept presented in \cite{Hubert05}. Originally, the orthogonal and score distances intend to identify outliers from one main group of observations. Therefore, robust estimators of location and scatter are used to estimate $\bs \mu$ and $\bs \Sigma_P$. Thus, the orthogonal and score distances are always interpreted with respect to the center and covariance structure  of the majority of observations. The larger those distances get, the less likely the evaluated observation belongs to the same group. While the original work is based on the assumption of one main group of observations and a small subset of outliers, we assume the presence of multiple groups. In the latter situation, robust estimators calculated from less than $50\%$ of the observations is not appropriate because in robust statistics a majority of observations has to be considered. Therefore, we alter the location and scatter estimates and estimate them from a small subset of observations where we try to select the observations from the same group. 

Since $SD_P(\bs x )$ and $OD_P(\bs x )$ are both measures for similarity with respect to a location and covariance matrix, we define 
\begin{align}
OSD_P(\bs x) &= f(SD_P(\bs x), OD_P(\bs x)) & \bs x \in \mb{R}^p, \\
&f: \mathbb{R}^2 \rightarrow \mathbb{R} \nonumber \\
&f \text{ monotonically increasing in } OD_P \text{ and } SD_P \nonumber
\end{align}
as a new univariate measure for similarity, always to be interpreted in reference to a location, a covariance matrix, and the dimensionality $q$ of the projection space, which in case of \cite{Hubert05} is given by the number of components used for the robust principal component analysis. Examples for such functions $f$ are provided in \cite{Pomerantsev2008}.

We utilize a subclass of the presented projections defined by $\mathbb{P}$. 
Let $\mcI$ denote a set of $q$ indices $\mcI_1, \dots, \mcI_q$ of $\bs X$, $\mcI \in \mc{P}(1,\dots, n): |\mcI| = q$, where $\mc{P}$ is the power set. $\bs X_{\mcI}$ defines the matrix of scaled and centred selected observations. To scale and centre the observations, we use a location estimator
\begin{equation}
\bs{\hat{\mu}}_\mcI = \bar{\bs{x} }_{\mcI } = \frac{1}{q} \sum_{i\in \mcI} \bs{x}_i 
\label{muhat}
\end{equation}
and a scale estimator
\begin{align}
\bs{\hat{ \sigma}}_\mcI &= (\sqrt{Var(x_{\mcI_1 1}, \dots, x_{\mcI_q 1} )}, \dots, \sqrt{Var(x_{\mcI_1 p}, \dots, x_{\mcI_q p} )})' \\
&= (\hat{\sigma}_{\mcI 1}, \dots, \hat{\sigma}_{\mcI p})' \nonumber ,
\end{align}
where $\bs{x}_{\mcI_k} = (x_{\mcI_k 1}, \dots, x_{\mcI_k p})'$ denotes the $k$-th selected observation and $Var$ is the empirical variance. $\bs x^c_{\mcI_k}$ denotes the centred observation $\bs{x}_{\mcI_k}$: 
\begin{align}
\bs x^c_{\mcI_k} &= \bs x_{\mcI_k} - \bs{\hat{\mu}} = (x^c_{\mcI_k 1}, \dots x^c_{\mcI_k p} )' \\
\bs X_{\mcI} &= \left(
\left( \frac{x_{{\mcI_1}1}^c}{\hat \sigma_{\mcI 1}}, \dots, \frac{x_{{\mcI_1}p}^c}{\hat \sigma_{\mcI p}} \right)',
\dots,
\left( \frac{x_{{\mcI_q}1}^c}{\hat \sigma_{\mcI 1}}, \dots, \frac{x_{{\mcI_q}p}^c}{\hat \sigma_{\mcI p}} \right)'
 \right)' \label{XI} 
\end{align}
The matrix $\bs{X_\mcI}$ can be represented via a singular value decomposition:
\begin{equation}
\bs X_{\mathcal{I}} = \bs U_\mathcal{I} \bs D_\mathcal{I}  \bs V_\mathcal{I}' \label{svd}
\end{equation}

Note that the centring of the observations reduces the rank of the data matrix by one. Therefore, under the assumption of $q < p$, the rank of $\bs V_\mcI'$, which is equal to the rank of $\bs X_\mcI$, is $q-1$. This assumption is reasonable due to the focus on high-dimensional data. If $q<p$ does not hold, the dimension of the space is small enough such that a data transformation is not required.
$\bs V_\mcI'$ from the decomposition in Equation \eqref{svd} provides a projection matrix onto the space spanned by the $q$ observations selected in $\mcI$. $\bs V_\mcI'$ represents an element of $\mathbb{P}$ since the dimension of the projection space is equal to the rank of $\bs V_\mcI'$ which is  $q-1$. For such a projection, we can measure the similarity of any observation from $\mathbb{R}^p$ to the selected observations using the location estimation from Equation \eqref{muhat} and covariance matrix describing the covariance structure in the projection space, provided by the selection itself as follows:
\begin{align}
\bs {\hat{\Sigma}}_\mcI &= \frac{1}{q-1} (\bs V_\mcI \bs X'_\mcI )(\bs V_\mcI \bs X'_\mcI)' \label{sigmahat}
\end{align}
Using the provided definitions and notation, we can define a univariate measure $OSD_\mcI(\bs x)$ for similarity  between an observation $\bs{x} \in \mb{R}^p$ and a set of observations, defined by $\mcI$:
\begin{align}
OSD_\mcI(\bs x) & = f(SD_\mcI(\bs x), OD_\mcI(\bs x)), & \bs x \in \mb{R}^p \label{osd}\\
SD_\mcI (\bs x) &= \sqrt{ (\bs V_\mcI'(\bs x-{ \bs {\hat{\mu}}_\mcI}))' \bs{ \hat{\Sigma}}_\mcI^{-1} (\bs V_\mcI'(\bs x-{\bs{\hat{\mu}}_\mcI}))}. \label{sdi} \\
OD_\mcI (\bs x) &= || \bs x - \bs{\hat{\mu}}_\mcI - \bs V_\mcI \bs V_\mcI'(\bs x-\bs{\hat{\mu}}_\mcI) ||
\label{odi}\end{align}

\subsection{Guided projections algorithm  }

To create a sequence of non-random projections, we aim to identify a set of $q$ observations, project all observations onto the space spanned by those $q$ observations, and use $OSD_\mcI$ to measure the similarity between an observation $\bs x \in \mathbb{R}^p$ and the selected group of observations. 
In general, $q$ is a configuration parameter which needs to be adjusted based on the data set to be analysed. Depending on both the expected number of observations in groups in the data structure and on the sparsity of the data set, we typically select $q$ between 10 and 25.
Out of the selected group of observations,
we replace one observation after another by a new observation and therefore get a new projection space leading to new measures for similarity.

To identify a set q of starting observations, we exploit the Euclidean distances between all observations. Let $d_{ij}$ denote the Euclidean distance $d(\bs{x}_i,\bs{x}_j) = || \bs{x}_i - \pmb{x}_j ||$ between observation $\bs{x}_i$ and $\bs{x}_j$. $d_{i(k)}$ denotes the $k^{th}$ smallest distance from $\bs{x}_i$:
\begin{equation}
\min_{j \in \{ 1, \dots,n \}} d_{ij} = d_{i(1)} \leq \dots \leq d_{i(n)} = \max_{j \in \{ 1, \dots,n \}} d_{ij}
\end{equation}

Similar to the k-nearest-neighbor approach \citep[e.g.][]{altman1992}, we identify a dense group of $q$ observations given by their indices {$\mcI^0_1, \dots, \mcI^0_q$. Let $i_0 = \arg \min\limits_{i \in \{ 1, \dots,n \}} d_{i(q)}$ denote the index of the observation with the smallest distance to the $q^{th}$-closest observation and $\bs X_{\mcI^0}$ the centered and scaled matrix of observations as defined in Equation \eqref{XI}:
\begin{equation}
\mcI^0 = \{ \mcI_1^0, \dots , \mcI_q^0 \} \label{I0} = \{ j: d_{i_0j} \leq d_{i_0(q)} \} 
\end{equation}

Note that in Equation \eqref{I0} we assume that the number of observations in $\mcI^0$ is equal to $q$ even though the second equality does not hold in general. In the case of ties, more than $q$ observations may fulfill the criterion $d_{i_0j} \leq d_{i_0(q)}$ of Equation \eqref{I0}. In such a case, we randomly select from the tied observations to be added to $\mcI^o$, such that $q$ observations are selected.

During the determination of the sequence of projections, we always add the observation with the smallest $OSD$ to the set of selected observations. To keep the dimensionality of the projected space constant, which ensures comparability of $OSD$s, we remove one observation each time we add an observation.  Assuming the observations are ordered in a certain sense, each observation remains in the group of selected observations for $q$ projections before it is removed again. 

To identify the observation $\bs{x}_{i_1}$ to be added in the first step, we solely need to consider $OSD_{\mcI^0}$ defined in Equation \eqref{osd}. The set of observations available to be selected is defined by $A^0$:
\begin{align}
A^0 &= \{ 1, \dots, n \} \backslash \mcI^0 \label{def:A0} \\
i_1& = \arg \min_{i \in A^0} OSD_{\mcI^0}(\bs{x}_i) 
\end{align}

To identify the observation to be removed, we need to provide an order of $\mcI^0$ first, which is determined by using leave-one-out distances $(LOD)$. Sorting all elements from $\mcI^0$ decreasingly according to $LOD$ provides the sorted starting observations and the first selected observation $i_1$ defined by $I^1$:
\begin{align}
LOD_{\mcI^0} (j,i_1) &= OSD_{\{ \mcI^0 \backslash \{ j \} \} \cup \{ i_1 \}}(\pmb{x}_j) & \forall j \in \mcI^0 \\
I^1 &=  (j_1, \dots , j_q, i_1) =(\iota_1^1, \dots, \iota^1_{q+1}) & j_k \in \mcI^0, k = 1,\dots,q   \label{I1}\\
& LOD_{\mcI^0}(j_1, i_1) \geq \dots \geq LOD_{\mcI^0}(j_q, i_1) \nonumber \\
A^1 &= A^0 \backslash i_1 = \{ 1, \dots, n \} \backslash I^1 \\
\mcI^1 &= \{ \mcI^0  \backslash j_1 \} \cup \{ i_1 \} 
\end{align}

$\mcI^1$ and $A^1$ again denote the index sets of observations selected in the first step and the remaining observations available for selection after the first step, respectively. After this first step, for any following step, in general for the $s^{th}$ step, two projections, represented by $\mcI_L$ and $\mcI_R$ are relevant for selecting a new observation:
\begin{align}
\mcI_L &= \{  \iota_1^{1}, \dots \iota_{q-1}^{1} \} \\
\mcI_R &= \{  \iota_2^{1}, \dots \iota_{q}^{1} \} 
\end{align}
The notation $L$ and $R$ comes from the left and right end of the series of indexes in $I^1$ representing the first and the last $q$ observations.

The reason to consider multiple projections is based on the assumption that we start from a dense region of the data distribution. By adding one observation we move away from this dense region in one direction. Once the observations at the border of this direction have been reached, the remaining observations are far away from the selection, yet close to the initially selected observations in the center. Figure \ref{fig:2directions} visualises this issue.

\ifplots
\begin{figure}[!h]
\centering
\begin{overpic}[width=\linewidth]{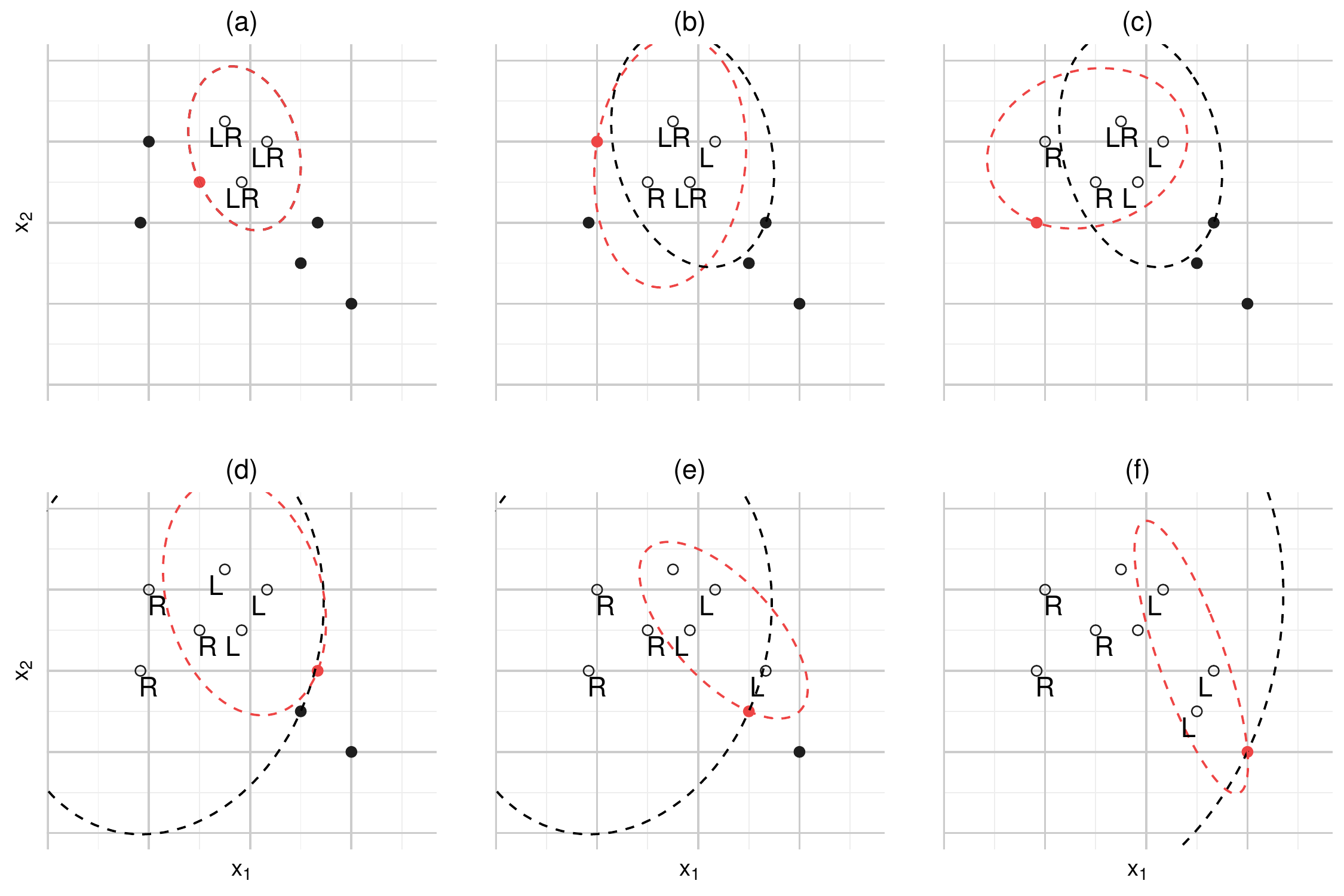}
\put(70,250){\red{$i_1$}}
\put(110,290){$OSD_{\mcI^0}$}
%step2
\put(207,283){\red{$i_R$}}
\put(195,228){$OSD_{\mcI_{R}}$}
\put(285,295){$OSD_{\mcI_{L}}$}
\put(202,176){$i_L$}
%step3
\put(370,240){\red{$i_R$}}
\put(455,245){$i_L$}
%step4
\put(115,60){$i_R$}
\put(125,82){\red{$i_L$}}
%step5
\put(260,60){$\red{i_L}=i_R$}
%step6
\put(450,46){$\red{i_L}=i_R$}
\end{overpic}
\caption{Visualisation of the selection procedure. To keep the observations in a constant location for each plot we use a two-dimensional space. The distances $OSD_\mcI$ to a selection of observations $\mcI$ are represented by dashed ellipses. The red ellipse represents the smaller distance and therefore the choice for the next observation to be selected. If an observation is part of 
$\mcI_L$ or $\mcI_R$ is marked with an L or R respectively. Filled points represent observations which have not been selected so far, empty circles have been selected before or are part of a current selection. The next observation to be added to the sequence is marked by a red dot.
}
\label{fig:2directions}
\end{figure}
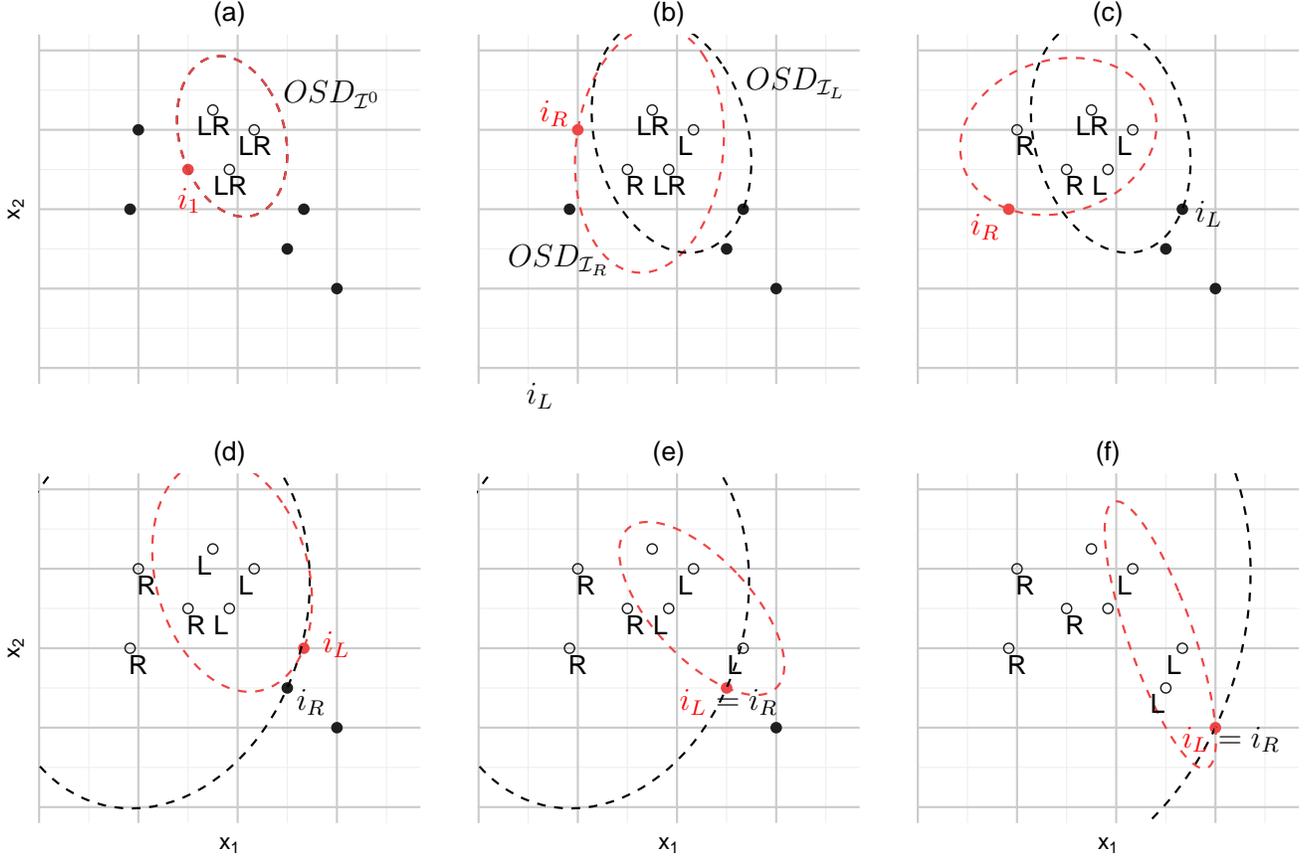
\fi

Since we aim at a series of projections as consistent as possible, we always select the projection with the smallest distance. In the showcase in Figure \ref{fig:2directions} we show the selection of $\mcI^0$ and the first observation $i_1$ in plot (a). Plot (b) to (f) represent the steps $1$ to $5$ of our procedure. The two ellipses represent the $OSD$, based on $\mcI_L$ and $\mcI_R$ respectively. The choice of observation to be added is marked as a red dot. Starting from plot (d) we notice that the selection $\mcI_R$, represented by the observations marked with an R, requires a large $OSD_{\mcI_R}$ to add an additional observation. Therefore, starting from (d) we add observations to the left end of the series $I^s$. In general it makes sense to consider all previous projections. However, to create a series of projections where we can look for structural changes and visualize a development, we limit ourselves to $\mcI_L$ and $\mcI_R$.

Depending on the smallest $OSD$ to either $\mcI_L$ or $\mcI_R$, the newly added observation, the new set of sorted observations $I^s$, and the new set of available observations for future projections $A^s$ are determined for the $s^{th}$ step, provided $s \geq 2$ holds:
\begin{align}
i_L &= \arg \min_{i \in A^{s-1}} OSD_{\mcI_L}(\bs{x}_i) \\
i_R &= \arg \min_{i \in A^{s-1}} OSD_{\mcI_R}(\bs{x}_i) \\
I^s &= \left\{
    \begin{matrix}
      ( i_L, \iota_1^{s-1},\dots,\iota^{s-1}_{s-1+q}),   \\ 
      ( \iota_1^{s-1},\dots,\iota^{s-1}_{s-1+q}, i_R), 
    \end{matrix} \right. &
     \begin{matrix}
      OSD_{\mcI_L}(\bs{x}_{i_L}) \leq  OSD_{\mcI_R}(\bs{x}_{i_R})  \\ 
      \mbox{else}
    \end{matrix} \\
   & = (\iota^s_1, \dots, \iota^s_{s+q}) \nonumber \\
 A^s &= \{ 1, \dots, n \} \backslash I^s 
\end{align}

$I^s$ is a superset of $I^{s-1}$ for all $s \geq 1$ and provides all information about the sequence of previous projections. In total, there are $n-q+1$ projections available which are determined after $n-q$ steps. Therefore, we can define the guided projections $GP$ based on $I^{n-q}$ alone.
\begin{align}
GP(\bs{x}) = &(GP_1(\bs{x}) , \dots, GP_{n-q+1}(\bs{x})) \\
GP_j(\bs{x}) &= OSD_{\{  \iota^{n-q}_j, \dots, \iota^{n-q}_{j+q-1}   \} }(\bs{x}) & j \in 1, \dots, n-q+1 \label{GPj}
\end{align}

As a result, we receive one series of measures for each observation. Whenever the measure is small, the observation is likely from the same group as the respective selected observations. Thus, structures in data can be identified by looking for similar behaviour in $GP(\bs x)$.

\subsection{Additional insight on guided projections}
\label{subsec:insight}

\hspace{10pt}
\textbf{Choice for $\mf{OSD}$:} 
A variety of useful $OSD$s can be defined for guided projections. Some possibilities to combine orthogonal and score distances to a univariate measure are presented in \cite{Pomerantsev2008}.
The best choice for OSD depends on the distribution of the data structure. 
When dealing with high-dimensional data, especially sparse data where groups are best described by different variables, the orthogonal distance contributes more to the group separation than the score distance. When dealing with low-dimensional data, the opposite is true.
Therefore, the decision on the most appropriate OSD needs to be met for each analysis individually  depending on the underlying data characteristics. Given the fact, that we deal with high-dimensional data and for reasons of simplicity we restrict the choice of OSD for this work to the orthogonal distance, utilizing the properties of the complement of the projection space which is often ignored \citep[e.g.][]{Gattone2012, Ilies2010}.
%Depending on the knowledge about the data structure, a proper choice for $OSD$ has to be made individually. 
\begin{equation}
OSD_{\mcI}(\bs x) = OD_\mcI(\bs x)
\label{usedosd}
\end{equation}

\textbf{Two-dimensional visualisation of guided projections:} 
Each projection results in a representation of all observations by orthogonal and score distances which can be visualised in a two-dimensional plane.
The series of projections $GP(\bs x)=(GP_1(\bs x), \dots, GP_{n-q+1}(\bs x))$ typically starts with observations from one group. Therefore, the observations to be selected in the following steps are observations which are similar to the selected observations and thus likely from the same group. By replacing only one observation per projection, we achieve a high correlation between $OSD$s created by consecutive projections. Each step represents a slight rotation of the two-dimensional $OD$-$SD$-plane, the observations are  projected onto. This behaviour is represented in Figure \ref{fig:seriesofprojections} where the projection space is always spanned by 10 observations. 

\ifplots
\begin{figure}[bht]
\centering
\includegraphics[width=\linewidth]{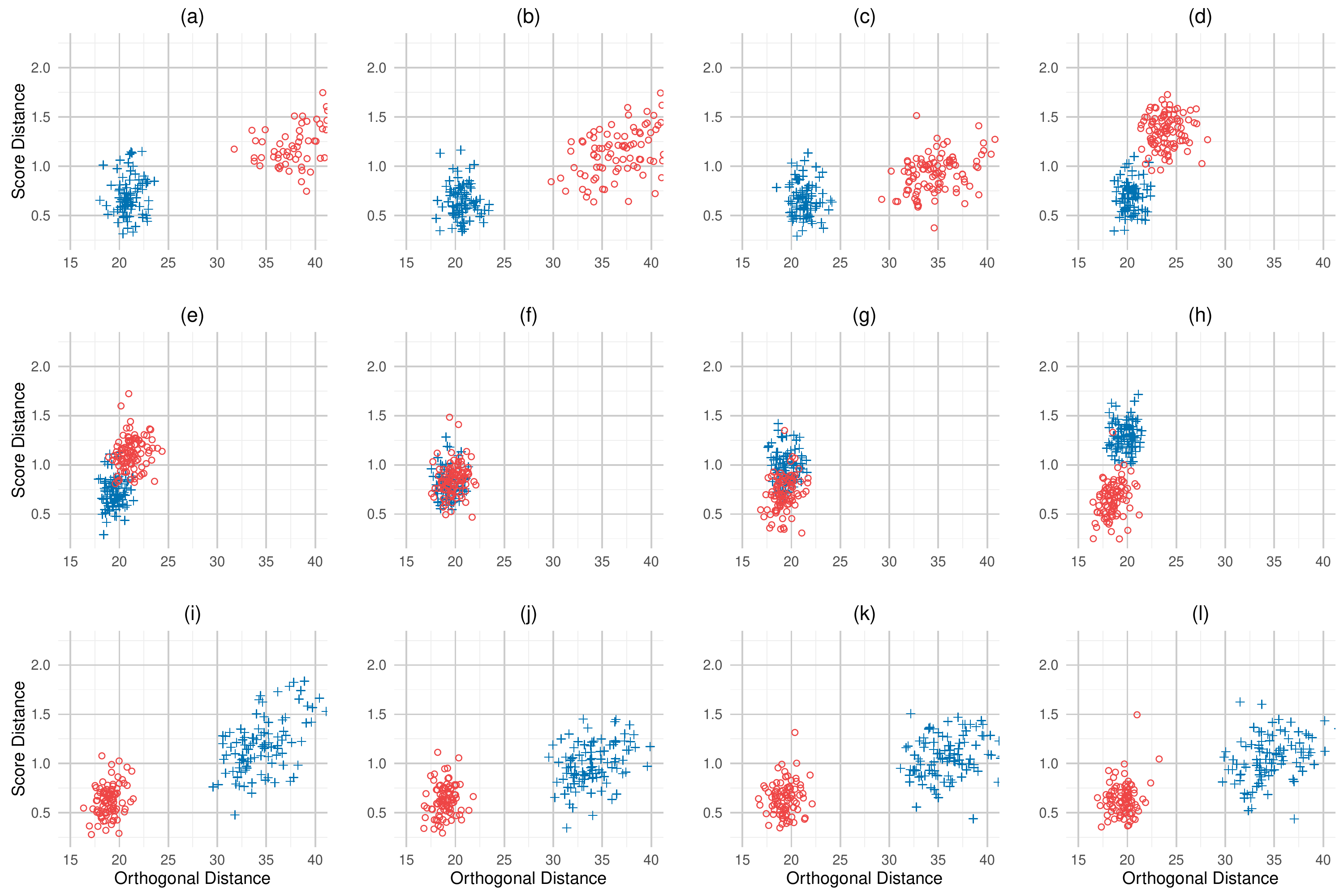}
\caption{Subset of the series of projections for simulated data,
consisting of two groups with 100 observations each, generated
from two different fifty-dimensional normal distributions. The groups are visualised with red circles and blue plus symbols. Each plot represents one step of guided projections, where all observations are projected onto the space spanned by 10 selected observations.}
\label{fig:seriesofprojections}
\end{figure}
\fi

In Figure \ref{fig:seriesofprojections}, the plots (a) to (d) show projections where all selected observations are taken from the blue (plus symbols) group. Figure (e) shows the first time where an observation from the red (circles) group is selected. Therefore, the distances for the red group start decreasing. In plot (g) the majority of selected observations is taken from the red group. In plot (h) only one blue observation remains in the selection. Starting from plot (i) in the third row, the groups are separated again since all observations for the projection are selected from the red group.

\textbf{Specific behaviour of OD and SD for guided projections:} 
Assume one of the projection matrices $\bs V_{\mcI^s}$, where $\mcI^s$ represents the selected observations in the $s^{th}$ step. 
Let us consider plot (a) of Figure \ref{fig:seriesofprojections} as an example. One could argue that critical values can be directly provided separating the red from the blue group for this projection, making the rest of the sequence obsolete. Details for the determination of those critical values for orthogonal distances and score distances are provided in \cite{QuadraticForms1992} and \cite{Pomerantsev2008}. The problem with this argument can be described as follows.

The possibility of separating two or more groups is based on the assumption that all selected observations are taken from the same group and an estimation of location and the covariance matrix based on this group only can be provided.  Therefore, such a decision needs to be made after the initial selection. Thus, only $q$ observations are available for the required estimation of location and covariance in the $q-1$ dimensional space. This estimation cannot be provided due to the following properties for all $s \in \{ 0,\dots ,n-q+1 \}$:
\begin{equation}
OD_{\mcI^s}(\bs x) = 0,  \hspace{10pt} \iff  \hspace{10pt} \bs x \in span(\{\bs x_i:i \in \mcI^s \})  \label{odxi}  
\end{equation}
\begin{equation}
  SD_{\mcI^s}(\bs x_i)  = \frac{q-1}{\sqrt{q}}, \hspace{15pt} \forall i \in \mcI^s \text{ and } q=|\mcI^s |, s \in \{ 0,\dots ,n-q+1 \}    \label{sdxi}
\end{equation}
The proof of these statements can be found in the Appendix.
Since there is no variation in the orthogonal and score distance for the selected observations for $\mcI^s$, the parameters for the critical values, which are based on the variation, cannot be derived. The orthogonal and score distances for observations of $\mcI^s$ are extremely distorted and do not follow the expected theoretical distribution of $OD_{\mcI^s}$ and $SD_{\mcI^s}$.

\subsection{Visualisation of guided projections}

Guided projections can be visualised in a diagnostic plot.
In such a plot, the series of $OSDs$ is shown for each observation.
As an example, consider the
data set used in Figure \ref{fig:seriesofprojections}. 
Due to Equation \eqref{odxi}, any selected observation will have an orthogonal distance of zero
for certain projections,
and therefore in our application an $OSD$ of zero, as defined in Equation \eqref{usedosd}.

Figure \ref{fig:diagnostic1} shows the change in $OSD$ by modifying the projection direction, which is achieved by substituting one observation in the selection spanning the projection space. Each observation is selected once. Therefore, for each projection, one observation drops to zero from a non-zero level and one observation goes up to a non-zero level. 

Given the 200 observations, selecting 10 observations for each projection results in a total number of 191 projections.
For the first 85 projections, all observations are selected from group one (blue dashed lines). During this procedure, no significant changes occur. Starting with the $86^{th}$ projection though, which is the same projection as plot (e) of Figure \ref{fig:seriesofprojections}, we see some mixed projections and a structural change in $OSD$ for both groups. The $OSDs$ of the observations from one group drop to a lower level while the $OSDs$ of the observations from the other group increase.

Such a structural change in guided projections clearly indicates the presence of a second group in the analysed data structure. In general, observations whose $OSD$ stays close to each other 
during the whole sequence of projections
are expected to belong to the same group. 

\ifplots
\begin{figure}[!htb]
\centering
\includegraphics[width=\linewidth]{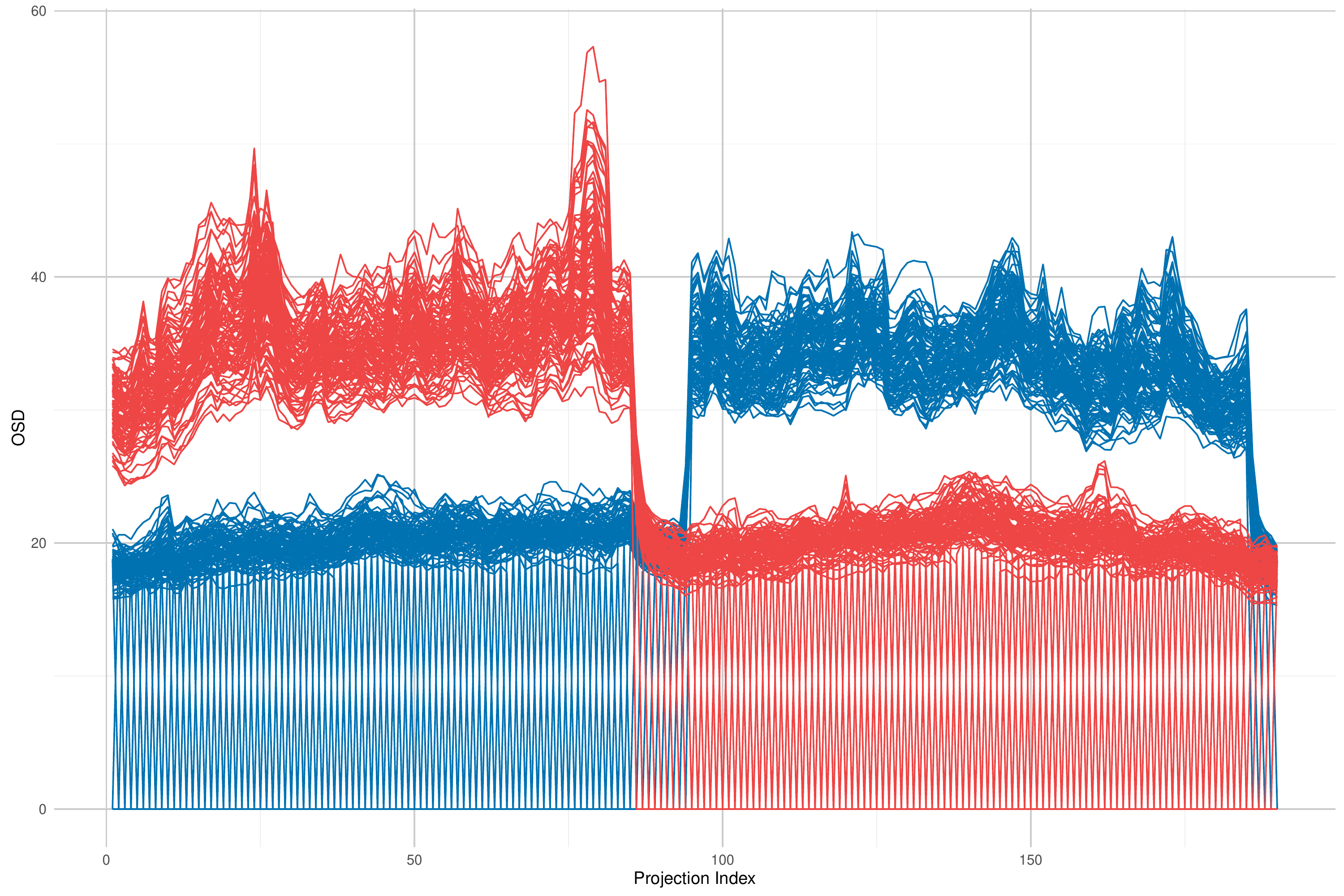}
\caption{Diagnostic plot utilizing guided projections for the simulated data from Figure \ref{fig:seriesofprojections}. The colors represent the two clusters, originally located in a fifty-dimensional space. The projection index on the x-axis stands for the index $j$ of $GP_j(\bs x)$ of Equation \eqref{GPj}. For each observation we can follow the change in $OSD$ while slightly changing the projection direction. Similar observations are represented in parallel lines, close to each other.} 
\label{fig:diagnostic1}
\end{figure}
\fi

%========================================================================
\section{Simulations}\label{sec:simulation}

The aim of this section is to measure the effect of data transformations on the separation of present groups in simulated data. We consider the data transformation approaches introduced in Section \ref{sec:motivation}: Classical PCA [PCA], Sparse PCA [SPC], Diffusion Maps [DIFF], and Random Projections [RP]. We use two simulated multivariate normally distributed data setups to measure the impact of noise variables as well as the impact of differences in covariance structures. 
The effects themselves are measured by a selection of common cluster validity measures. 

\subsection{Evaluation Measures}

An overview of internal evaluation indices is presented in \cite{Desgraupes2013}. All measures can be directly accessed through the R-package \textit{clusterCrit} \citep{clusterCrit}. The provided indices depend on various measures like total dispersion, 
within-group scatter and between-group scatter. Some of those measures heavily depend on the dimensionality of the transformation space. Thus, depending on the design of the validity measures, a lower dimensional space is often preferred over a high-dimensional space even though the quality of separation decreases with decreasing dimensionality. We use two simulations visualised in Figure \ref{fig:validitymeasures} to demonstrate this aspect. 
In the first setup we generate $k$ simulated independent normally distributed variables. Group one uses a mean value of $1$, while group two uses mean values of $-1$. The more variables are used, the better the expected separation should be. The second  simulation setup always uses 50 of those variables and in addition adds $k$ normally iid variables with mean value of zero for both groups. Those non-informative variables theoretically reduce the quality of the group separation. For a selection of popular validation measures we simulate those two setups, varying $k$ between $1$ and $350$. Note that not all original measures should be maximised. Therefore we transformed all measures which should be minimized, like the Banfeld Raftery index, in such a way that they are to be maximised to simplify Figure \ref{fig:validitymeasures}. 

\ifplots
\begin{figure}[tbh]
\centering
\includegraphics[width=\linewidth]{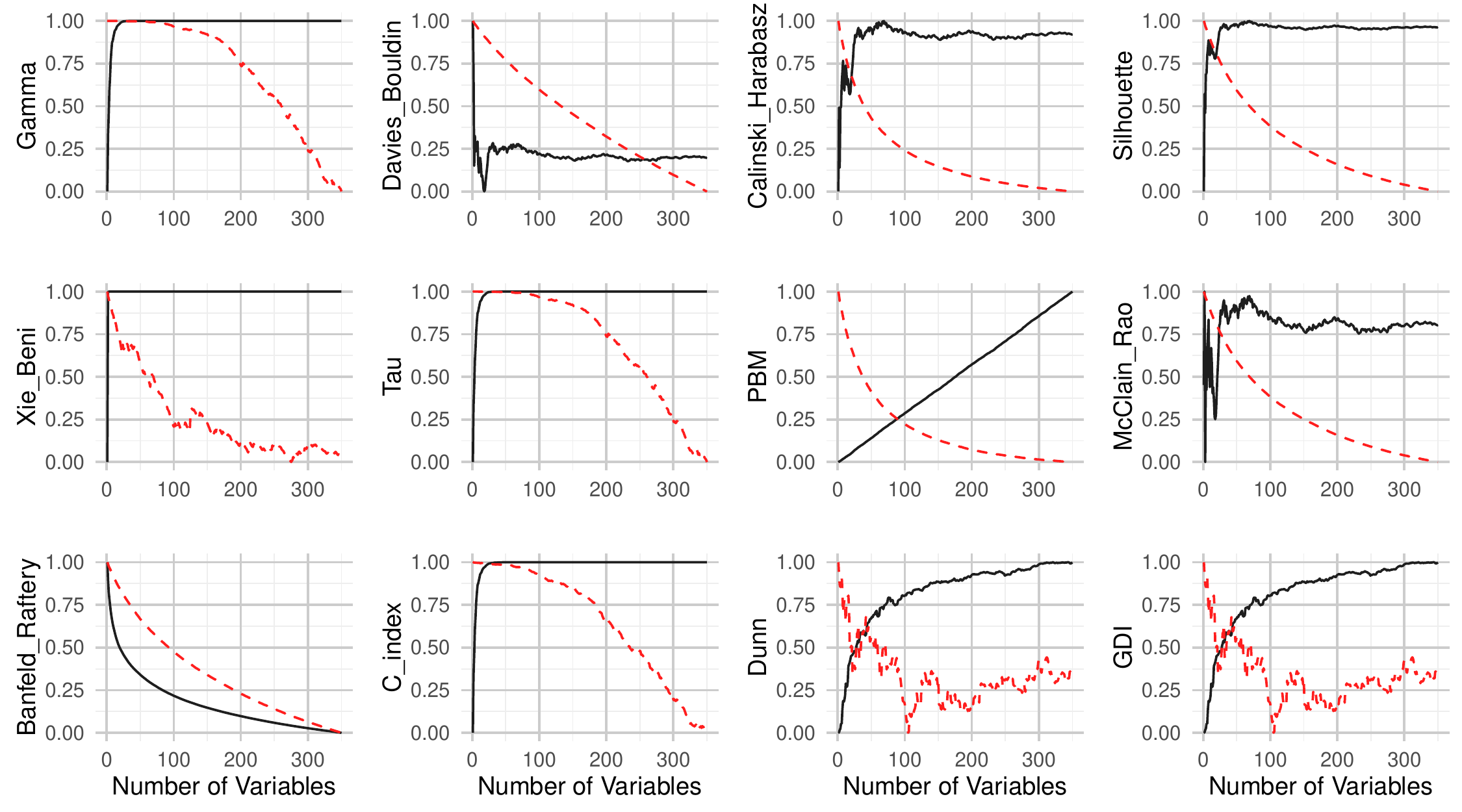}
\caption{The solid (black) line refers to the previously described setup one  (informative variables only), the dashed (red) line to setup two  (including non-informative variables). The transformed validity measures for both setups have been independently scaled to the interval $[ 0, 1 ]$ for a better visualisation. Both lines are depending on the number of variables related to the respective setup. In total, 1000 observations are simulated for each simulation setup and group to evaluate the considered measures.}
\label{fig:validitymeasures}
\end{figure}
\fi

The decision on which indices to consider for the evaluation is based on the simulation results. Validity measures with a non-monotonous development for the second setup (Xie Beni, Dunn Index and GDI) are excluded. Also measures with a decreasing development in the first setup (Davies Bouldin and Banfield Raftery) or a large fluctuation range in setup 1 (Calinski Harabasz and McClain Rao) have been excluded. Among the remaining validation measures, based on their popularity we decided to include the \textit{Gamma} index \citep{Baker1975}, the \textit{Silhouette} index \citep{Rousseeuw1987}, and the \textit{C index} \citep{Hubert1976} for the evaluation of 
the group structure of
data transformations.

In addition to the selected validity measures, we are interested in the effect of data 
transformations before applying clustering procedures. Therefore, we perform hierarchical Ward clustering \citep{Ward1963} after applying the data transformations and evaluate the clustering result using the \textit{F-measure} \citep{Larsen1999}. 

\subsection{Parameter optimisation}

A number of data transformations has been presented in Section \ref{sec:motivation}. Each of them is depending on one or more configuration parameters, leading to different 
quality of the projections and thus directly affecting the validation measures. 

All methods are optimised for each data set individually. For each parameter we set upper and lower boundaries in which we optimise the parameters for each specific data transformation method and validation measure. This way we make the methods comparable since a specific parameter set might work better for one transformation than for another providing an unfair advantage for one method. The same is true for specific validation measures. The optimisation itself is performed by allowing a discrete number of parameters within their boundaries and performing and evaluating each combination of parameters. Hereinafter we present parameters to be optimised for the compared data transformations.

\textbf{PCA}: 
For principal component analysis the only parameter that needs to be adjusted is the proportion of variance of $\bs X$ which should be represented in the projection space. This can be translated to the number of components considered to span the projection space.  This dimension is optimised for any number between $1$ and the rank of $\bs X$, which is the maximum number of possible components.

\textbf{SPC}: The considered sparse principal component analysis by \cite{Witten2009} uses two optimisation parameters. The first parameter is the number of sparse components, the second parameter the degree of sparsity defined by the sum of absolute values of elements of the first right singular vector of the data matrix. The number of components is optimised equivalently to PCA. The sparsity parameter is optimized between $1$ and the square root of the number of columns of the data as recommended in \cite{Witten2009}.

\textbf{DIFF}: 
Diffusion maps utilize an $\epsilon$-parameter to describe the degree of localness in the diffusion weight matrix. A recommended starting point is $2med_{knn}^2$, where $med_{knn}^2$ represents the squared median of the $k^{th}$ nearest neighbour. By varying $k$ between $0.5\%$ and $3.5\%$ of the number of observations, which extends the recommended $1\%$ to $2\%$, we adjust the $\epsilon$-parameter. The number of components to describe the transformation space is adjusted in the same way as for $PCA$.

\textbf{RP}: For random projections we repeatedly project the observations on a $k$ dimensional projection space 500 times. We optimise $k$ between $1$ and $k_{max}$. The upper limit $k_{max}$ is the maximum number of components available in PCA for real data and the number of informative variables for simulated data.

\textbf{GP}: For guided projections, only one parameter needs to be adjusted, namely the number of observations in each projection. We propose to optimise this number between $5$ and $30$.

While performing hierarchical clustering, the number of clusters emerges as an additional configuration parameter. To provide a fair comparison, we allow any possible number of clusters between 1 and the number of observations, and report the best possible result. Figure \ref{fig:example_optimisation} visualises the optimisation for the Gamma index and the F-measure for an exemplary data set for SPC. 

\ifplots
\begin{figure}[hbt]
\centering
\includegraphics[width=\linewidth]{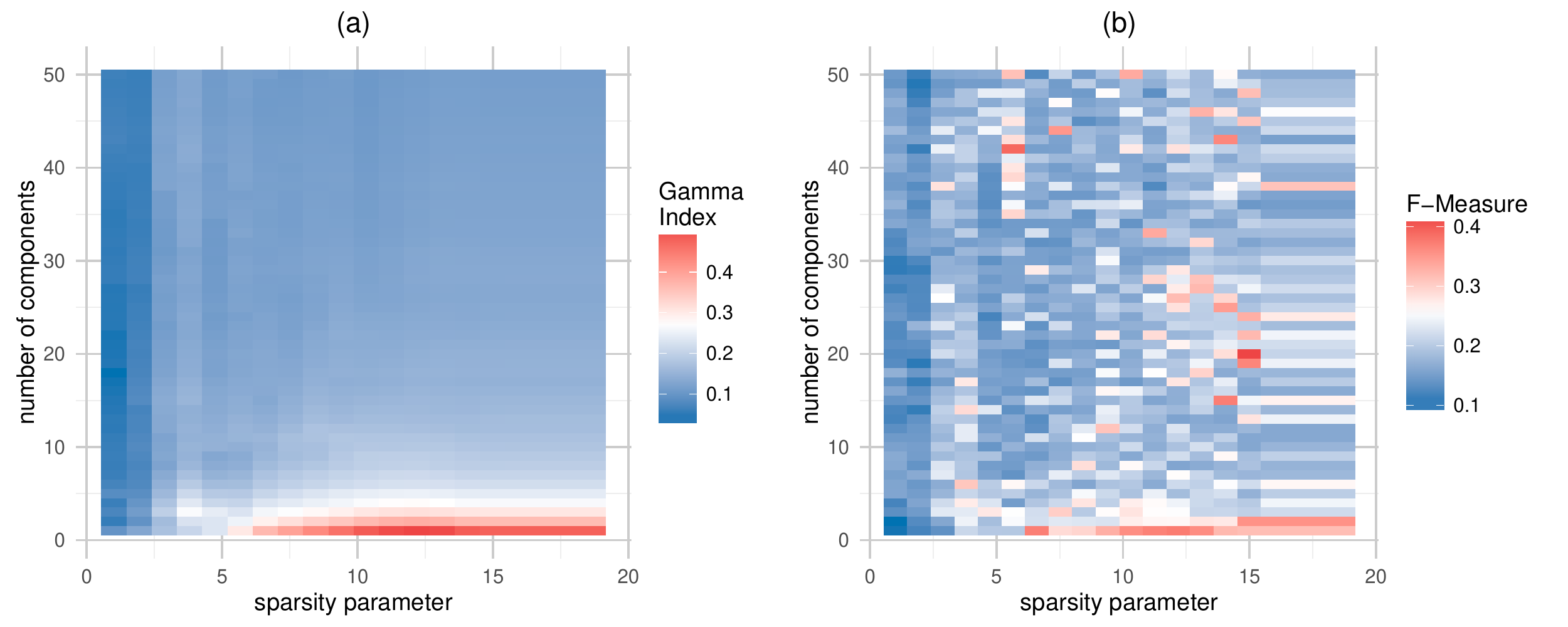}
\caption{The optimisation procedure for SPC is visualised. 
On the x-axis the sparsity parameter is presented, on the y-axis the number of sparse components.
The quality of each parameter combination is presented by the color of the respective combination. 
Red corresponds to a high value of the considered validity measure, blue to a low value.  Figure (a) shows the optimisation for the Gamma index, Figure (b) for the F-measure. For each index, the individual optimum is selected. The sparsity parameter for the F-measure is selected slightly larger than for the Gamma index. The optimal F-measure requires 20 sparse principal components while the Gamma index uses one.}
\label{fig:example_optimisation}
\end{figure}
\fi

Note that we do not compare with projection pursuit, since the aim of
this approach is to identify a low-dimensional projection (one to three dimensions) 
revealing the group structure of the data. 
We evaluated the final projection of a guided tour from \cite{Wickham2011} and found no significant difference to the performance of random projections. Such an evaluation is unfair though since two-dimensional projections are being compared with methods that incorporate multiple or higher dimensional projections. Therefore, projection pursuits are not considered for the full evaluation.

\subsection{First simulation setting}
\label{sec:data}

The first simulated data setup consists of two groups of observations, where the observations are drawn from different multivariate normal distributions $X_1 \sim N(\bs \mu_1, \bs \Sigma_1)$ and $X_2 \sim N(\bs \mu_2, \bs \Sigma_2)$. 
The parameters are as follows:
\begin{align}
\bs \mu_1 =& (\bs{0}_{50}, \bs{0.5}_{50}, \bs{0}_{250})'   \label{setup1_mu1}\\
\bs \mu_2 =& (\bs{0}_{r}, \bs{-0.5}_{50}, \bs{0}_{300-r})' \label{setup1_mean2}
\end{align}
\begin{equation}
\bs \Sigma_1 = \left( \begin{array}{ccc}
\bs I_{50} & 0   &  0\\
0 &\bs \Sigma^{rand_2}_{50} & 0 \\
0 & 0 & \bs I_{250} 
\end{array} \right)
\end{equation}
\begin{equation}
\bs \Sigma_2 = \left( \begin{array}{ccc}
\bs I_{r} & 0   &  0 \\
0 &\bs  \Sigma^{rand_2}_{50} & 0 \\
0 & 0 &\bs I_{300-r} 
\end{array} \right) 
 \label{setup1_cov2}
\end{equation}

In \eqref{setup1_mu1} to \eqref{setup1_cov2}, $\bs{0}_r$ and $\bs{0.5}_r$ denote a vector of length $r$ with $0$ or $0.5$ entries, respectively. $\bs I_r$ denotes an $r$-dimensional unit matrix and $\bs \Sigma_{50}^{rand_1}$ and $\bs \Sigma_{50}^{rand_2}$ represent randomly generated, fifty-dimensional covariance matrices.

By varying $r$ we modify the subspace where the informative variables are located. For $r=51$, a $50$ dimensional informative subspace is present but this subspace is informative for both present groups. For other values of $r$, the informative variables of $X_2$ are getting shifted away from the informative variables from $X_1$. An interesting aspect of this setup is the fact that the expected difference between the two groups changes with $r$. The expected distance between $X_1$ and $X_2$ is based on the number of informative variables as well as on the expected distance for each informative variables. In fact, the expected distances turn out to be
\begin{equation}
E(|| X_1-X_2 ||) = \sqrt{ 50-\frac{1}{2}min(50,|51-r|) } .
\label{exp_dist}
\end{equation}
This distance is maximised for $r=51$ and is decreasing with any changes in $r$ leading to the expectation of a maximised separation for $r=51$. For each $r$ between 1 and 100, we repeatedly simulate the setup 25 times. For each simulated data set we report optimised validation measures.

\ifplots
\begin{figure}[!htb]
\centering
\includegraphics[width=\linewidth]{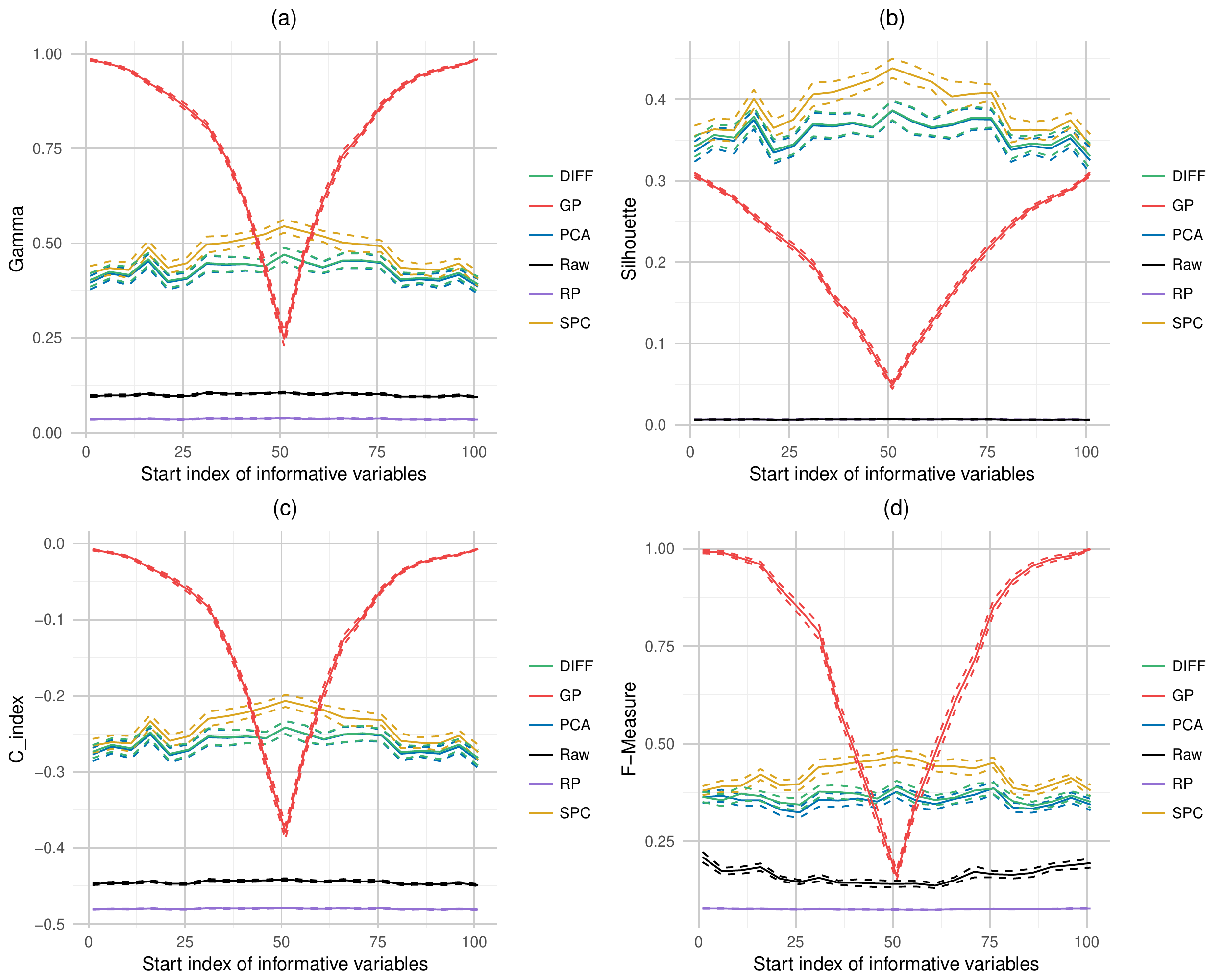}
\caption{For each selected validation measure, we show the mean performance (solid lines) of the 5 considered data transformations as well as their respective standard error (dashed line). The performance of no transformation is shown by the \textit{Raw} category. The start index of the informative variable on the x-axis refers to the parameter $r$ of Equation \eqref{setup1_mean2} and \eqref{setup1_cov2}. 
The results for DIFF and PCA are very similar and thus almost plot on top of
each other.}
\label{fig:indices_setup1}
\end{figure}
\fi

Each plot in Figure \ref{fig:indices_setup1} shows a similar individual behaviour for each method. 
The performance of principal component based methods increases with increasing expected distance between $X_1$ and $X_2$, which is described in Equation \eqref{exp_dist}, while the quality of guided projections increases with additional informative variables and especially with an increase in the shift of informative variables. This behaviour by guided projections occurs due to the following properties: When observations from the same group are selected, the subspace spanned by those observations describes the informative variables of those observations. Therefore, if the second group consists of different informative variables, the difference in orthogonal distances increases, which are used here for $OSD$. 
If the informative variables are the same though, the differences in the orthogonal space are expected to be the small. Since we completely ignore the score distances, guided projections are outperformed by principal component based methods in this case.
This feature is visible for all considered validation measures. Most validation
measures indicate that guided projections
clearly outperform the other projection methods if the number of informative (shifted)
variables increases. An exception is the Silhouette index, which declares guided projections as
the worst method. However, this might be quite specific in a two-group setting.

%\red{The comparison with projection pursuits are not included for the reason of an unfair comparison. Utilising the final two dimensional projection from guided tours \cite{Wickham2011} we note that the average performance of those projections is .}

\subsection{Second simulation setting}

The second simulated data setup uses three groups drawn from multivariate normally iid stochastic variables $X_1 \sim N(\bs \mu_1, \bs \Sigma_1)$, $X_2 \sim N(\bs \mu_2, \bs \Sigma_2)$ and  $X_3 \sim N(\bs \mu_3, \bs \Sigma_3)$ with the following parameters:
\begin{align}
\bs \mu_1 =& (\bs 1_{25},  \bs 1_{25}, \bs 0_{25}, \bs 0_r)' \label{setup2_mu1} \\
\bs \mu_2 =& (\bs 1_{25},  \bs 0_{25}, \bs 1_{25}, \bs 0_r)' \\
\bs \mu_3 =& (\bs 0_{25},  \bs 1_{25}, \bs 1_{25}, \bs 0_r)' 
\end{align}

\begin{equation}
\bs \Sigma_1 = \left( \begin{array}{cccc}
\bs \Sigma^{rand_{1,1}}_{25}  & \bs \Sigma^{rand_{1,2}}_{25}    &  0 & 0 \\
\bs \Sigma^{rand_{1,3}}_{25}  &\bs \Sigma^{rand_{1,4}}_{25} & 0 & 0 \\
0 & 0 & \bs I_{25} & 0 \\
0 & 0 & 0 &\bs I_{r}
\end{array} \right)
\label{sig1_setup2}
\end{equation}

\begin{equation}
\bs \Sigma_2 = \left( \begin{array}{cccc}
\bs \Sigma^{rand_{2,1}}_{25} & 0 & \bs \Sigma^{rand_{2,2}}_{25}    &  0  \\
0 & \bs I_{25} & 0 & 0 \\
\bs \Sigma^{rand_{2,3}}_{25} & 0 &\bs \Sigma^{rand_{2,4}}_{25} &  0 \\
0 & 0 & 0 &\bs I_{r}
\end{array} \right)
\label{sig2_setup2}
\end{equation}

\begin{equation}
\bs \Sigma_3 = \left( \begin{array}{cccc}
 \bs I_{25} & 0 & 0  & 0 \\
0 & \bs \Sigma^{rand_{3,1}}_{25}  & \bs \Sigma^{rand_{3,2}}_{25}    &  0 \\
0 &\bs \Sigma^{rand_{3,3}}_{25}  &\bs \Sigma^{rand_{3,4}}_{25} & 0 \\
0 & 0 & 0 &\bs I_{r}
\end{array} \right)
\label{sig3_setup2}
\end{equation}
Similar as before, $\bs 0_r$ and $\bs 1_r$ represent vectors of length $r$ with $0$ and $1$ entries, respectively. The matrices $\left( \begin{array}{cc} 
 \pmb \Sigma^{rand_{i,1}}_{25}  & \pmb \Sigma^{rand_{i,2}}_{25}    \\
\pmb \Sigma^{rand_{i,3}}_{25}  &\pmb \Sigma^{rand_{i,4}}_{25}
\end{array} \right)$ from Equation \eqref{sig1_setup2} to \eqref{sig3_setup2} represent randomly created 50 dimensional covariance matrices. Therefore, $\bs \Sigma_1$, $\bs \Sigma_2$ and $\bs \Sigma_3$ represent covariance matrices too. The first $75$ variables are informative variables, while the remaining $r$ variables are non-informative. With increasing $r$, the separation between the present groups gets increasingly masked. The focus of this simulation setup is the robustness of data transformations towards non-informative variables.

The parameter $r$ is varied between $0$ and $1250$ leading to a $75$ to $1325$ dimensional space. For each setup we compare three groups of 100 simulated observations per group. 25 repeated simulations are performed for each evaluated $r$ by randomly creating different covariance matrices. 
 
 \ifplots
 \begin{figure}[!htb]
\centering
\includegraphics[width=\linewidth]{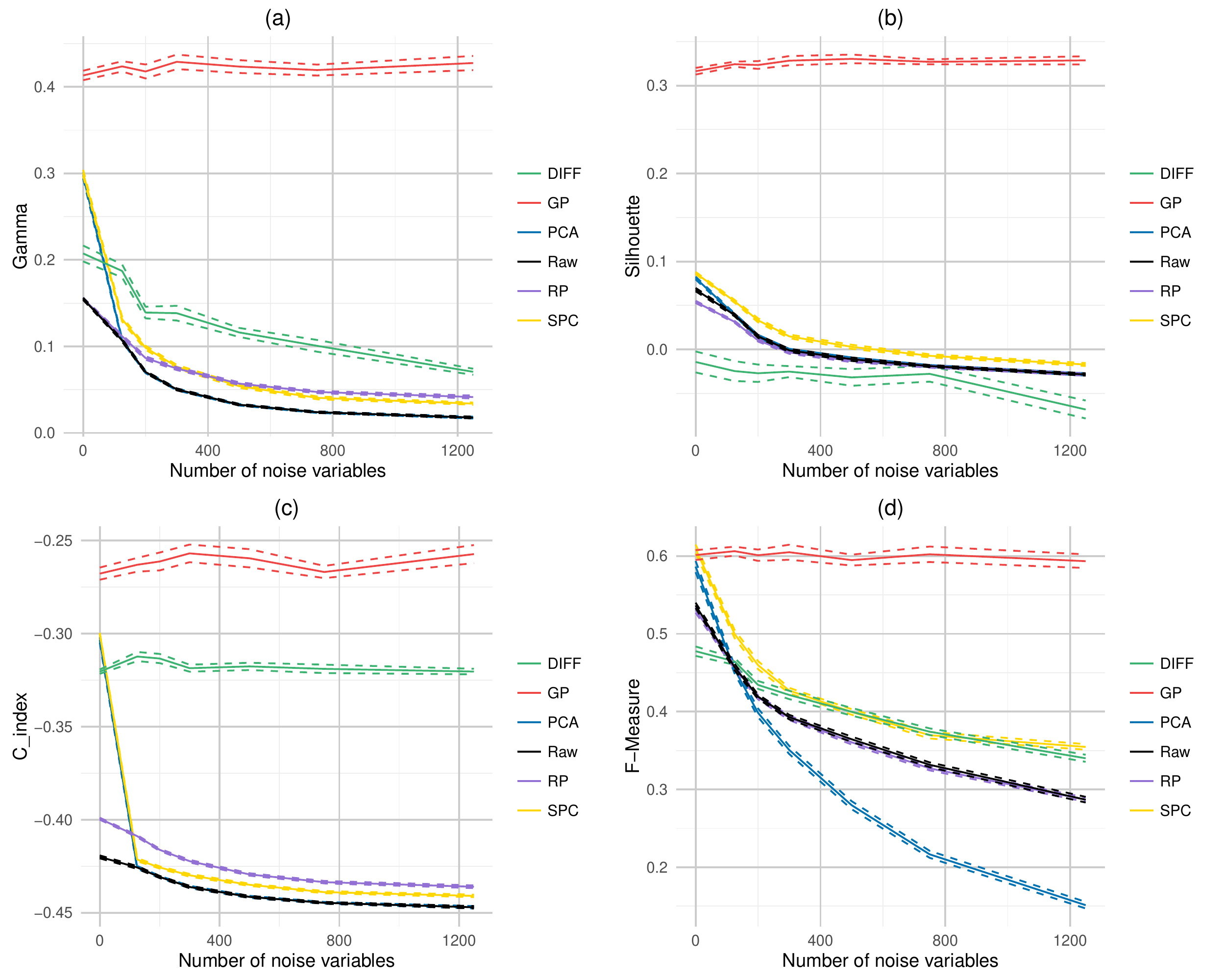}
\caption{For the selected validation indices, we analyse the impact of additional noise variables. The mean optimal performance and the respective standard error is visualised for an increasing number of noise variables by solid and dashed lines for each transformation. In general we expect a decrease in quality with increasing noise variables.}
\label{fig:indices_setup2}
\end{figure}
\fi
 
Figure \ref{fig:indices_setup2} shows the effect of increasing $r$ non-informative variables  on the quality of the considered data transformation, based on the different validation measures. The number of non-informative variables $r$ refers to $r$ in Equation \eqref{setup2_mu1} to \eqref{sig3_setup2}. For each method and validation measure but guided projections for all measures and diffusion maps for C-index index we see the quality of transformations being affected in the same way as the level of separation is affected for the untransformed data. For guided projections though, there seems to be no impact from additional non-informative variables. Compared to setup 1 where only two groups were present, guided projection clearly outperform all other transformation regardless of the validation index.

%=====================================================================
\section{Real-world data sets}\label{sec:fruit}

The first real-world dataset we take into consideration is the fruit data set which is often used to demonstrate the stability of robust statistical methods \citep[e.g.][]{Hubert2004}. It consists of 1095 observations of spectra of three different types of melon labelled with \textit{D}, \textit{M} and \textit{HA}, presented in a 256 dimensional space of wavelength. It is known that the groups consist of subgroups due to changed illumination systems and changed lamps while cultivating the plants. 
Since we do not have labels for the subgroups, we only consider the originally provided labels. For those labels we randomly select $100$ observations per group repeatedly 50 times. 

Figure \ref{fig:fruit1} evaluates the separation of groups based on the Gamma index, the Silhouette measure, the C-index and the F-measure. Guided projections clearly outperforms all other transformations as well as the untransformed data situation. Only when measured with the C-index, diffusion maps perform better than guided projections. For all other validation measures though, diffusion maps perform below average. 

In addition to showing that the presentation of the observations with guided projections leads to a better group separation, we can visualise the transformation using the diagnostic plot. Figure \ref{fig:diag_fruit} visualises the transformation for all available observations. First, a group of projections, supporting the separation between the red and the green group can clearly be seen in the second half of the projections. Second, we can see additional group structure in the red group and a small number of outliers for almost all projections. The presence of outliers and additional group structure for this data set is well known \citep[e.g.][]{Hubert2004}. These subgroups, however, are not documented, and therefore an evaluation of the additionally observed group structure is not possible. 

\ifplots
\begin{figure}[!htb]
\centering
\includegraphics[width=\linewidth]{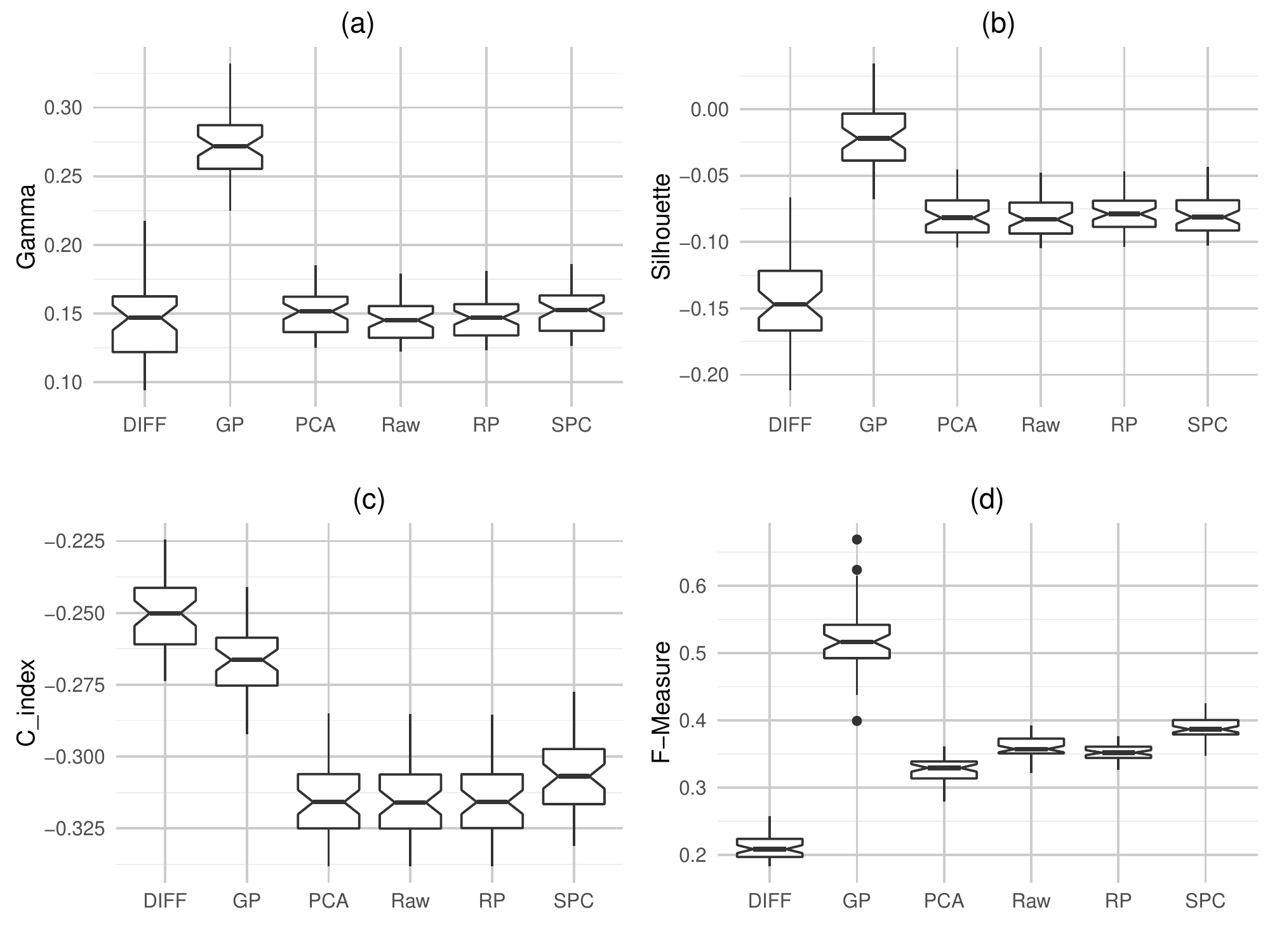}
\caption{The performance of data transformations is measured by four different validation measures. 50 randomly selected subsets of the fruit data set are evaluated, based on the originally provided labels.}
\label{fig:fruit1}
\end{figure}
\fi

\ifplots
\begin{figure}[!htb]
\centering
\includegraphics[width=\linewidth]{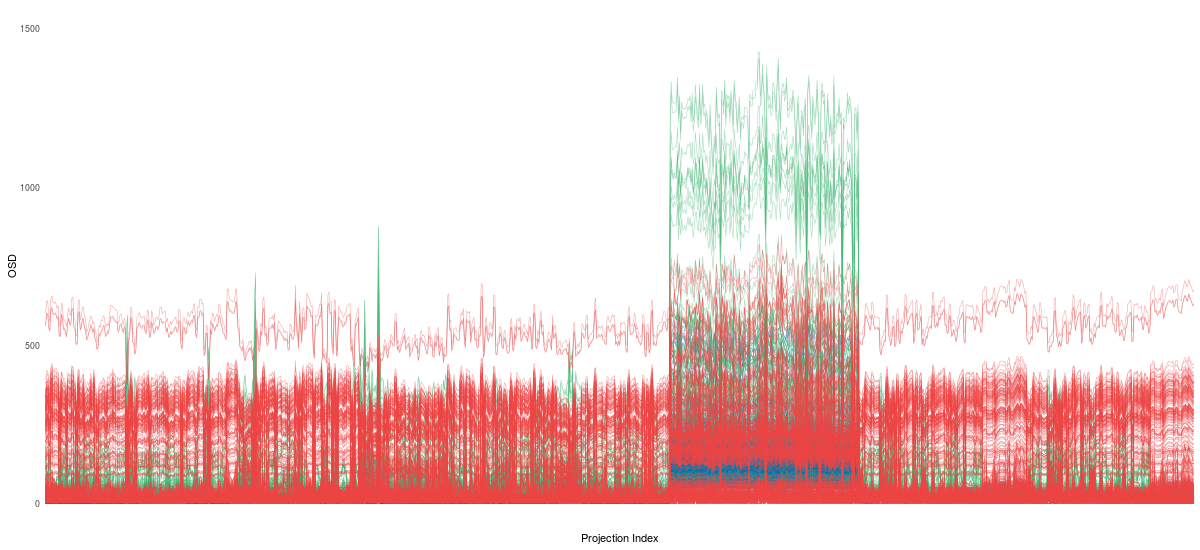}
\caption{Diagnostic plot for the full fruit data set. Three groups are present. Additional group structure can be adumbrated. Especially the presence of outliers is evident. The observed group structure reflects the changes in the illumination system while collecting data from melon growth as described in various publications \citep[e.g.][]{Hubert2004}.}
\label{fig:diag_fruit}
\end{figure}
\fi

To show that the identification of additional group structures and outliers can be achieved, utilizing diagnostic plots for guided projections we further introduce the \textit{glass vessels} \citep[e.g.][]{Filzmoser2008} dataset. Archaeological glass vessels 
from the $16^{th}$ and $17^{th}$ century were investigated by an electron-probe X-ray micro-analysis. In total, $1920$ characteristics are used to describe each vessel. 
The presence of outliers, especially in one out of the
four glass groups has been shown in previous studies \citep{Serneels2005}. We use the algorithm \textit{pcout} \citep{Filzmoser2008} to identify outliers in 
this group of observations. 
The diagnostic plot based on guided projections is visualised in Figure \ref{fig:glass_vessels}. We can see that the outliers from \textit{pcout}, drawn in red, correspond to the most 
remote observations
in the diagnostic plot. We can further identify additional group 
structure and some additional candidates for outliers. It is not clear, what underlying nature this group structure is identified from and it seems to be undocumented so far by statistical publications working on the very same glass vessels data set.
This information will be valuable for the analyst, because it can refer
to problems in the measurement process, or to inconsistencies in the observations which
are initially assumed to belong to one group.

\ifplots
\begin{figure}[!htb]
\centering
\includegraphics[width=\linewidth]{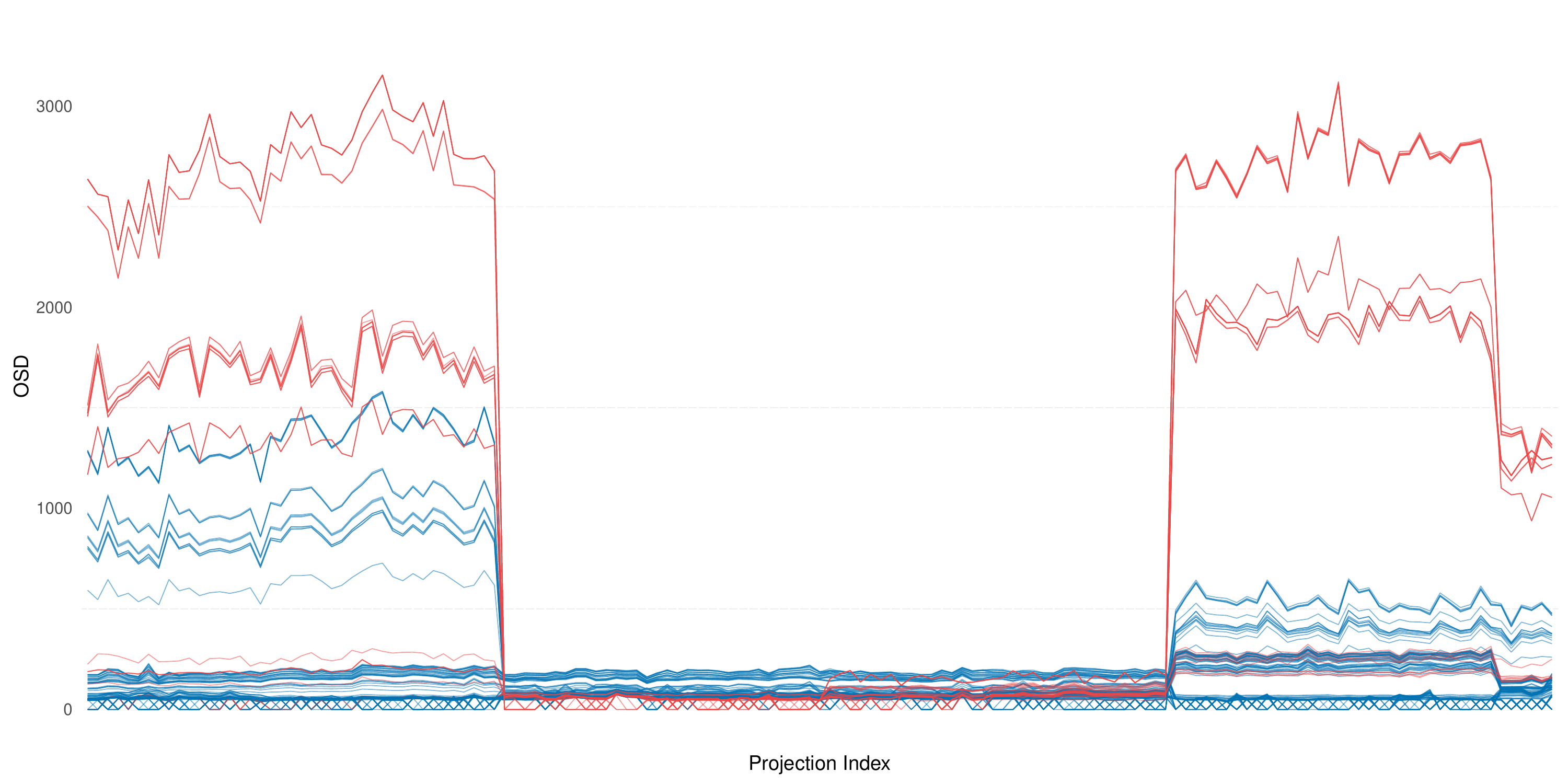}
\caption{Diagnostic plot for the glass vessel data set. Only the main group of glass vessels is considered. Red  lines correspond to identified outliers by the \textit{pcout} algorithm from \cite{Filzmoser2008}. }
\label{fig:glass_vessels}
\end{figure}
\fi

%====================
\section{Conclusions and outlook}\label{sec:conclusion}

We have proposed guided projections as an alternative to existing data transformations which are applied prior to data structure evaluation methods. 
We project all observations on the space spanned by a small number of $q$ observations which are selected in a way such that they are likely to come from the same group. We then exchange observations in this selection one by one and therefore create a series of projections. Each projection can then be treated as a new variable, but only the complete series is used for investigating the grouping
structure contained in the data.
Note that this approach differs conceptually from projection pursuit
approaches, where the focus is on identifying one (or several) low-dimensional
projections of the data that reveal the group structure.

While guided projections is motivated by the separation of groups using the full available information, its application can be extended onto all types of data structure analysis which is affected by high-dimensionality like outlier detection, cluster analysis, or discriminant analysis. Furthermore, a way for identifying the existence of group structure is provided by the introduced visualisation of guided projections. This concept can be further extended to new diagnostic plots for identifying outliers and group structures in the data.

The results based on simulated data show the advantages and limitations of guided projections in comparison to other data transformation methods. Given favourable conditions in the data structure, namely informative variables in different subspaces, guided projections can vastly improve the degree of separation between existing groups in the data. Furthermore, guided projections turned out to be a lot more robust against additional non-informative variables. The results based on the real world data sets also prove the practical importance of guided projections. 

There are multiple ways to further improve the concepts of guided projections. First, we can remove the restriction of considering only the projections $\mcI^s_L$ and $\mcI^s_R$ for each step. Instead, we can consider every projection of previous steps. Removing this limitation allows a more complex network of projections instead of an ordered series of projections. 
The setup requires additional research. 
The second adjustment is the implementation of different distance measures in the projection space. While PCA-based transformations create an orthogonal basis in the projection space, guided projections are highly correlated. Only few projections often provide enough information for a perfect separation. Identifying these projections is a task of its own.

Furthermore, a detailed evaluation of possible measures for $OSD$ needs to be performed to allow a proper evaluation of the limitations and possibilities of guided projections.

% ===============================================================
\section*{Appendix}\label{sec:appendix}

\appendix
Equation \eqref{odxi} and \eqref{sdxi} can be proven using the decomposition $\bs x = \bs z_1 + \bs z_2$, where $\bs z_1 \in span(\{\bs x_i:i \in \mcI^s \})$ and $\bs z_2 \in span^{\bot}(\{\bs x_i:i \in \mcI^s \})$. $span$ represents all possible linear combinations of its observations and $span^{\bot}$ its orthogonal complement. Specifically, we write $\bs z_1 = \sum\limits_{i \in \mcI^s} a_i \bs x_i$. For the equality of Equation~\eqref{odxi} it is important to note that also $\bs{\hat{\mu}}$ is a linear combination of $\bs x_i, i \in \mcI^s$, with constant coefficients $\frac{1}{q}$. Thus, we can use the property $\bs x_i = \bs V_{\mcI^s} \bs D_{\mcI^s} \bs u_i$, which holds for all $i \in \mcI^s$ where $\bs u_i$ represents the respective right singular vector:
\begin{align}
OD_{\mcI^s} (\bs z_1)  = || \bs z_1 - \bs{\hat \mu} - \bs V_{\mcI^s} \bs V'_{\mcI^s}(  \sum\limits_{i \in \mcI^s}a_i\bs x_i - \sum\limits_{i \in \mcI^s}\frac{1}{q}\bs x_i )||, \hspace{12pt} a_i \in \mathbb{R} \hspace{5pt} \forall i \in \mcI^s \nonumber \\
=|| \bs z_1 - \bs{\hat \mu} - (  \sum\limits_{i =1}^q a_i  \bs V_{\mcI^s} \bs V'_{\mcI^s} \bs V_{\mcI^s} \bs D_{\mcI^s} \bs u_i - \sum\limits_{i =1}^q\frac{1}{q} \bs V_{\mcI^s} \bs V'_{\mcI^s} \bs V_{\mcI^s} \bs D_{\mcI^s} \bs u_i )|| \label{eqpr1}
\end{align}

Since $\bs V'_{\mcI^s} \bs V_{\mcI^s} = \bs I$, one can see that the two linear combinations in Equation \eqref{eqpr1} sum up to $\bs z_1$ and $\bs{\hat{\mu}}$ respectively. Therefore, Equation \eqref{eqpr1} can be simplified to
\begin{equation}
OD_{\mcI^s}(\bs z_1)  = || \bs z_1 - \bs{\hat\mu} - (\bs z_1 - \bs{\hat\mu})|| = 0,
\end{equation}
which proves Equation \eqref{odxi}. To show Equation \eqref{sdxi} we first note that $\bs{ \hat \Sigma}_{\mcI^s}$ can be written as $\frac{1}{q-1} \bs D^2_{\mcI^s}$ and due to Equation \eqref{svd} $\bs V'_{\mcI^s} \bs x_i = \bs D_{\mcI^s} \bs u_i$ holds. Therefore, we can rewrite the squared score distances for $\bs x_i$ for all $i \in \mcI^s$ as:
\begin{align}
&SD^2(\bs x_i) = \left(\bs V'_{\mcI^s} (\bs x_i-\bs{\hat{\mu}}) \right)' \bs{ \hat \Sigma}^{-1}_{\mcI^s} \left(\bs V'_{\mcI^s} (\bs x_i-\bs{\hat{\mu}}) \right) \\
&= \left( \bs D_{\mcI^s} \bs u_i - \frac{1}{q} \sum\limits_{j \in \mcI^s} \bs D_{\mcI^s} \bs u_j \right)'
(q-1) \bs D_{\mcI^s}^{-2} 
\left( \bs D_{\mcI^s} \bs u_i - \frac{1}{q} \sum\limits_{l \in \mcI^s} \bs D_{\mcI^s} \bs u_l \right) \nonumber \\
&= (q-1) \left( \bs u'_i \bs u_i -
 \frac{1}{q}\sum\limits_{j\in \mcI^s} \bs u'_j \bs u_i
 - \frac{1}{q} \bs u'_i  \sum\limits_{l\in \mcI^s} \bs u_l +
\frac{1}{q^2}\left(\sum\limits_{j\in \mcI^s} \bs u'_j\right)
\left(\sum\limits_{l\in \mcI^s} \bs u_l\right)  \right). \nonumber
\end{align}

Due to $\bs U_{\mcI^s}$ being a unitary matrix and therefore $\bs u'_i \bs u_j = \delta_{ij}$, $\delta_{ij}$ denoting Kronecker's delta, this expression can be simplified. 
\begin{equation}
SD^2(\bs x_i) = (q-1)\left(1-\frac{1}{q}-\frac{1}{q}+\frac{q}{q^2}\right)=\frac{(q-1)^2}{q}
\end{equation} 
which proves Equation \eqref{sdxi}.

% ===============================================================


\bigskip

\begin{thebibliography}{}

\bibitem[Abdi and William, 2010]{Abdi2010}
Abdi, H. and William, L. (2010).
\newblock Principal component analysis.
\newblock {\em Computational Statistics}, pages 443--459.

\bibitem[Achlioptas, 2003]{Achlioptas2003}
Achlioptas, D. (2003).
\newblock Database-friendly random projections: Johnson-lindenstrauss with
  binary coins.
\newblock {\em Journal of Computer and System Sciences}, 66(4):671--687.

\bibitem[Altman, 1992]{altman1992}
Altman, N. (1992).
\newblock An introduction to kernel and nearest-neighbor nonparametric
  regression".
\newblock {\em The American Statistician}, 46:175--185.

\bibitem[Baker and Hubert, 1975]{Baker1975}
Baker, F.~B. and Hubert, L.~J. (1975).
\newblock Measuring the power of hierarchical cluster analysis.
\newblock {\em Journal of the American Statistical Association},
  70(349):31--38.

\bibitem[Coifman and Lafon, 2006]{Coifman2006}
Coifman, R.~R. and Lafon, S. (2006).
\newblock Diffusion maps.
\newblock {\em Applied and computational harmonic analysis}, 21(1):5--30.

\bibitem[Cook et~al., 1993]{Cook1993}
Cook, D., Buja, A., and Cabrera, J. (1993).
\newblock Projection pursuit indexes based on orthonormal function expansions.
\newblock {\em Journal of Computational and Graphical Statistics},
  2(3):225--250.

\bibitem[Cook et~al., 1995]{Cook1995}
Cook, D., Buja, A., Cabrera, J., and Hurley, C. (1995).
\newblock Grand tour and projection pursuit.
\newblock {\em Journal of Computational and Graphical Statistics},
  4(3):155--172.

\bibitem[De~Leeuw, 2011]{Leeuw2011}
De~Leeuw, J. (2011).
\newblock History of nonlinear principal component analysis.
\newblock {\em in Visualization and Verbalization of Data}.

\bibitem[Desgraupes, 2013]{Desgraupes2013}
Desgraupes, B. (2013).
\newblock Clustering indices.
\newblock {\em University of Paris Ouest-Lab Modal'X}, 1:34.

\bibitem[Desgraupes, 2016]{clusterCrit}
Desgraupes, B. (2016).
\newblock {\em clusterCrit: Compute clustering validation indices.}
\newblock R package version 1.2.7.

\bibitem[Donoho et~al., 2000]{Donoho2000}
Donoho, D.~L. et~al. (2000).
\newblock High-dimensional data analysis: The curses and blessings of
  dimensionality.
\newblock {\em AMS Math Challenges Lecture}, pages 1--32.

\bibitem[Filzmoser et~al., 2008]{Filzmoser2008}
Filzmoser, P., Maronna, R., and Werner, M. (2008).
\newblock Outlier identification in high dimensions.
\newblock {\em Computational Statistics \& Data Analysis}, 52(3):1694--1711.

\bibitem[Friedman and Tukey, 1974]{Friedman1974}
Friedman, J.~H. and Tukey, J.~W. (1974).
\newblock A projection pursuit algorithm for exploratory data analysis.
\newblock {\em IEEE Transactions on Computers}, c-23(9):881--890.

\bibitem[Gattone and Rocci, 2012]{Gattone2012}
Gattone, S.~A. and Rocci, R. (2012).
\newblock Clustering curves on a reduced subspace.
\newblock {\em Journal of Computational and Graphical Statistics},
  21(2):361--379.

\bibitem[Gorban et~al., 2008]{Gorban2008}
Gorban, A.~N., K{\'e}gl, B., Wunsch, D.~C., Zinovyev, A.~Y., et~al. (2008).
\newblock {\em Principal manifolds for data visualization and dimension
  reduction}, volume~58.
\newblock Springer.

\bibitem[Guyon and Elisseeff, 2003]{Guyon2003}
Guyon, I. and Elisseeff, A. (2003).
\newblock An introduction to variable and feature selection.
\newblock {\em Journal of Machine Learning Research}, pages 1157--1182.

\bibitem[Hubert and Schultz, 1976]{Hubert1976}
Hubert, L. and Schultz, J. (1976).
\newblock Quadratic assignment as a general data analysis strategy.
\newblock {\em British journal of mathematical and statistical psychology},
  29(2):190--241.

\bibitem[Hubert et~al., 2005]{Hubert05}
Hubert, M., Rousseeuw, P., and Vanden~Branden, K. (2005).
\newblock Robpca: A new approach to robust principal component analysis.
\newblock {\em Technometrics}, 47:64--79.

\bibitem[Hubert and Van~Driessen, 2004]{Hubert2004}
Hubert, M. and Van~Driessen, K. (2004).
\newblock Fast and robust discriminant analysis.
\newblock {\em Computational Statistics \& Data Analysis}, 45(2):301--320.

\bibitem[Hung and Tseng, 2003]{Hung2003}
Hung, Y.-C. and Tseng, N.-F. (2003).
\newblock Extracting informative variables in the validation of two-group
  causal relationship.
\newblock {\em Computational Statistics}, pages 1151--1167.

\bibitem[Ilies and Wilhelm, 2010]{Ilies2010}
Ilies, I. and Wilhelm, A. (2010).
\newblock Projection-based partitioning for large, high-dimensional datasets.
\newblock {\em Journal of Computational and Graphical Statistics},
  19(2):474--492.

\bibitem[Larsen and Aone, 1999]{Larsen1999}
Larsen, B. and Aone, C. (1999).
\newblock Fast and effective text mining using linear-time document clustering.
\newblock In {\em Proceedings of the fifth ACM SIGKDD international conference
  on Knowledge discovery and data mining}, pages 16--22. ACM.

\bibitem[Lee and Cook, 2010]{Lee2010}
Lee, E. and Cook, D. (2010).
\newblock A projection pursuit index for large p small n data.
\newblock {\em Statistics and Computing}, 10(3):381--392.

\bibitem[Li et~al., 2006]{Li2006}
Li, P., Hastie, T.~J., and Church, K.~W. (2006).
\newblock Very sparse random projections.
\newblock In {\em Proceedings of the 12th ACM SIGKDD international conference
  on Knowledge discovery and data mining}, pages 287--296. ACM.

\bibitem[Mathai and Provost, 1992]{QuadraticForms1992}
Mathai, A. and Provost, S.~B. (1992).
\newblock {\em Quadratic forms in random variables: theory and applications}.
\newblock Marcel Dekker, Inc., New York.

\bibitem[Pomerantsev, 2008]{Pomerantsev2008}
Pomerantsev, A.~L. (2008).
\newblock Acceptance areas for multivariate classification derived by
  projection methods.
\newblock {\em Journal of Chemometrics}, 22:601--609.

\bibitem[Rousseeuw, 1987]{Rousseeuw1987}
Rousseeuw, P.~J. (1987).
\newblock Silhouettes: a graphical aid to the interpretation and validation of
  cluster analysis.
\newblock {\em Journal of computational and applied mathematics}, 20:53--65.

\bibitem[Serneels et~al., 2005]{Serneels2005}
Serneels, S., Croux, C., Filzmoser, P., and Van~Espen, P.~J. (2005).
\newblock Partial robust m-regression.
\newblock {\em Chemometrics and Intelligent Laboratory Systems}, 79(1):55--64.

\bibitem[Ward~Jr, 1963]{Ward1963}
Ward~Jr, J.~H. (1963).
\newblock Hierarchical grouping to optimize an objective function.
\newblock {\em Journal of the American statistical association},
  58(301):236--244.

\bibitem[Wickham et~al., 2011]{Wickham2011}
Wickham, H., Cook, D., Hofmann, H., Buja, A., et~al. (2011).
\newblock tourr: An r package for exploring multivariate data with projections.
\newblock {\em Journal of Statistical Software}, 40(2):1--18.

\bibitem[Witten et~al., 2009]{Witten2009}
Witten, D.~M., Tibshirani, R., and Hastie, T. (2009).
\newblock A penalized matrix decomposition, with applications to sparse
  principal components and canonical correlation analysis.
\newblock {\em Biostatistics}, page kxp008.

\bibitem[Zou and Hastie, 2005]{Zou2005}
Zou, H. and Hastie, T. (2005).
\newblock Regularization and variable selection via the elastic net.
\newblock {\em Journal of the Royal Statistical Society: Series B (Statistical
  Methodology)}, 67(2):301--320.

\bibitem[Zou et~al., 2006]{Zou2006}
Zou, H., Hastie, T., and Tibshirani, R. (2006).
\newblock Sparse principal component analysis.
\newblock {\em Journal of computational and graphical statistics},
  15(2):265--286.

\end{thebibliography}
\end{document}